\documentclass[12pt,prd,onecolumn,floatfix,nofootinbib]{revtex4}

\usepackage{amssymb}
\usepackage{amsmath}
\usepackage{graphicx,color,dcolumn,booktabs,bm}
\usepackage{longtable,lscape}
\usepackage{txfonts}
\usepackage{overpic}
\usepackage{indentfirst}
\usepackage{cases}
\usepackage{multirow}
\usepackage{epstopdf}
\usepackage{ulem}
\usepackage[colorlinks,
            citecolor=blue,
            anchorcolor=red,
            menucolor=red,
            linkcolor=red,
            filecolor=red,
            runcolor=red,
            urlcolor=blue,
            frenchlinks=false]{hyperref}
\usepackage{threeparttable}
\allowdisplaybreaks
\begin{document}

\title{Systematics of doubly heavy tetraquarks: nonstrange and strange}
\author{Kai-Kai Zhang$^{1}$}
\email{zhangkk314@outlook.com}
\author{Wen-Xuan Zhang$^{1}$}
\email{zhangwx89@outlook.com}
\author{Duojie Jia$^{2,3}$}\thanks{Corresponding author}
\email{djjia@qhit.edu.cn}
\affiliation{
	$^1$Institute of Theoretical Physics, College of Physics and
	Electronic Engineering, Northwest Normal University, Lanzhou 730070, China \\
	$^2$General Education Center, Qinghai Institute of Technology, Xining, 810000, China\\
	$^3$Lanzhou Center for Theoretical Physics, Lanzhou University, Lanzhou, 730000, China \\
	}
\date{\today}

\begin{abstract}
Masses, magnetic moments and color-spin structures of nonstrange and strange tetraquarks with two heavy quarks are systematically studied in QCD string picture with chromomagnetic interaction. Our mass computations combined with weak and radiative decays indicate that there are two doubly-heavy tetraquarks, the bottom-charmed tetraquark $T_{cb}(7173,01^+)^0$ and the doubly-bottom tetraquark $T_{bb}(10406,01^+)^-$, to be stable against the strong decay, having the lifetimes of $0.069$ fs and $326$ fs, respectively. Magnetic moments are also computed and found to be sensitive to chromomagnetic mixing of the color-spin configurations of these tetraquarks, being valuable probe to experimentally detect the nontrivial color configuration $6_{c}\otimes\bar{6}_{c}$.
\end{abstract}

\maketitle
\date{\today }

\section{Introduction}

Early at the birth of the quark model \cite{Gell-Mann:1964ewy,Zweig:1964ruk}, exotic hadron states like tetraquarks (with quark configuration $q^{2}\bar{q}^{2}$)and pentaquarks ($q^{4}\bar{q}$) had been suggested, which are also allowed in principle by the quantum chromodynamics (QCD) established later. Yet, such an exotic state has not been observed until 2003 when the Belle reported discovery of the first exotic hadron $X(3872)$ \cite{Belle:2003nnu}. Since then, many exotic candidates have been observed subsequently, which include the charmonium-like states the $Z_{c}(3900)$ \cite{BESIII:2013ris,Belle:2013yex}, the $Z_c(4020)$ \cite{BESIII:2013ouc}, the $X(4050)$ \cite{Belle:2008qeq} and pentaquark states \cite{LHCb:2015yax,LHCb:2019kea,LHCb:2020jpq,LHCb:2022ogu}. The first doubly-charmed tetraquak $T_{cc}(3875)^{+}$($IJ^{P}=01^{+}$) was discovered in 2021 by the LHCb collaboration \cite{LHCb:2021vvq,LHCb:2021auc}, which triggered considerable efforts of experimental searches and theoretical studies of multiquark states and other exotic hadrons \cite{Liu:2019zoy,Chen:2022asf,Liu:2024uxn}.

The multiquarks were explored theoretically in the 1970s based on the MIT bag model \cite{Jaffe:1976ig,Jaffe:1976ih}. Till now, multiquark states, mainly the heavy tetraquarks, have been discussed extensively based on pictures of hadronic molecular \cite{Liu:2019stu,Dong:2021bvy,Chen:2021vhg,Yan:2021wdl,Ren:2021dsi,Feijoo:2021ppq,Meng:2021jnw,Abolnikov:2024key,Gil-Dominguez:2024zmr,Wu:2024zbx}, the compact tetraquark \cite{Weng:2021hje,Guo:2021yws,Zhang:2021yul,Deng:2021gnb,Lin:2024gcm,Dong:2024upa}, and other explanations. For the static properties of exotic hadrons, much effort has been made using various approaches, including the constituent quark model \cite{Zhou:2022gra, Wang:2023bek, Lai:2024jfe, Lei:2023ttd}, the Bag model \cite{Zhang:2021yul, Yan:2023lvm, Zhang:2023hmg, Zhu:2023lbx, Zhang:2023teh}, the QCD sum rule \cite{Lee:2011nq, Aliev:2014pfa, Azizi:2021aib, Ozdem:2023rkx, Ozdem:2024lpk}, and the relativistic three-quark model \cite{Faessler:2006ft}, and so on. See Refs. \cite{Jaffe:2004ph,Chen:2022asf} for reviews.

In order to explore static properties of doubly heavy tetraquarks with and without strange quarks in the QCD string picture with chromomagnetic interaction, we extract the model parameters from the observed mass data of the light and heavy mesons, and utilize a simple mass relations with interquark binding $B_{ij}$ and flux-tube energy attached to each quark, which we proposed previously in Ref. \cite{Liu:2023vrk}, to compute the ground-state masses of the doubly heavy tetraquarks and chromomagnetic mixing of their color-spin structures. Further computation are given to magnetic moments of these doubly heavy tetraquarks related to their color-spin structures and the relative decay widths of the strongly stable tetraquarks.

After introduction, in Sect. \ref{sec:for}, a chromomagnetic interaction model in the framework of QCD string picture is utilized to give formulism of mass and magnetic moments of a DH tetraquark and the parameters in it are determined from the established ordinary hadrons. In Sect. \ref{sec:CD}, the masses and magnetic moments of nonstrange and strange DH tetraquarks $T_{cc}$, $T_{bb}$ and $T_{cb}$ are computed, and the different decay behaviors of them are discussed. We end with
conclusions and remarks in Sect. \ref{sec:cr}.

\section{Masses, magnetic moments and parameters}
\label{sec:for}

\subsection{Chromomagnetic interactions}
\label{sec:ci}

For ground state of the tetraquarks, the chromomagnetic interaction (CMI) model is given by the Sakharov-Zeldovich formula \cite{Keren-Zur:2007ytk, Liu:2023vrk}
\begin{equation}\label{for:Ha}
	\begin{aligned}
		M=M_{\text{CE}}+\left\langle H_{\text{CMI}}\right\rangle,M_{\text{CE}}=\sum\nolimits_i(m_i+E_i)+B,
	\end{aligned}
\end{equation}
where $M_{\text{CE}}$ denotes the chromo-electric energy, $H_{\text{CMI}}$ is the CMI energy, $m_i$ the mass of the $i$th quark or antiquark, $E_i$ the effective energy of the color flux attached to the $i$th (anti)quark.
Following Ref. \cite{Liu:2023vrk,Karliner:2014gca}, we introduce an enhanced binding energy, $B = \sum\nolimits_{i<j}B_{Q_iQ_j(\bar{Q}_j)}$, which arises from short-distance chromo-electric interactions between heavy quarks and heavy (anti)quarks or between heavy quarks and strange (anti)quarks \cite{Karliner:2014gca, Karliner:2017elp, Karliner:2018bms}.

For the quark pairs $cc$, $bb$ and $cb$ in the color antitriplet $\bar{3}_c$, we shall apply a simple relations in Ref. \cite{Liu:2023vrk} to determine the corresponding binding energies $B_{cc}$, $B_{bb}$ and $B_{cb}$ in the doubly heavy (DH) tetraquarks in the color rep. of $3_{c}\otimes\bar{3}_{c}$. For the quark pairs $cc$, $bb$ and $cb$ in the color anti-sextet $\bar{6}_c$, we employ the scaling of color factors of the rep. $6_c$ relative to the rep. $\bar{3}_c$, to evaluate the corresponding binding energies $B_{cc}$, $B_{bb}$ and $B_{cb}$ in DH tetraquarks in the color rep. of $6_{c}\otimes\bar{6}_{c}$. Another interquark term in Eq. (\ref{for:Ha}) is that of the hyperfine chromomagnetic interaction, which is of spin and color dependent \cite{DeRujula:1975qlm}
\begin{equation}\label{for:cmi}
	\begin{aligned}
		H_{\text{CMI}} &=-\sum_{i<j}(\bm{\sigma}_i\cdot\bm{\sigma}_j)(\bm{\lambda}_i\cdot\bm{\lambda}_j)\frac{A_{ij}}{m_im_j},\\
	\end{aligned}
\end{equation}
where $\bm{\sigma}_i$ are the Pauli matrices for quark spin and $\bm{\lambda}_i$ are the Gell-Mann matrices for quark colors. The effective couplings $A_{ij}$ denote the strength of the CMI between the $i$th and $j$th (anti)quarks. The color and spin factors for the quark pair $ij$ in a DH tetraquark are defined as the respective matrix element of $\bm{\sigma}_i\cdot\bm{\sigma}_j$ and $\bm{\lambda}_i\cdot\bm{\lambda}_j$, and can be evaluated via the following formulas \cite{Luo:2017eub, Zhang:2021yul}:
\begin{equation}\label{factors}
	\begin{aligned}
		\langle\bm{\lambda}_i\cdot\bm{\lambda}_j\rangle_{mn}&=\sum_{\alpha=1}^{8}\text{Tr}\left(c^\dagger_{im}\lambda^\alpha c_{in}\right)\text{Tr}\left(c^\dagger_{jm}\lambda^\alpha c_{jn}\right),\\
		\langle\bm{\sigma}_i\cdot\bm{\sigma}_j\rangle_{xy}&=\sum_{\alpha=1}^{3}\text{Tr}\left(\chi^\dagger_{ix}\sigma^\alpha \chi_{iy}\right)\text{Tr}\left(\chi^\dagger_{jx}\sigma^\alpha \chi_{jy}\right),\\
	\end{aligned}
\end{equation}
where $n$ and $m$ denote the quark color and $x$ and $y$ the quark spin.

For strong decay of an exotic DH tetraquark, which were discussed in Refs. \cite{Weng:2020jao,An:2020vku,Zhang:2023teh} among others, we shall apply the S-wave two-body decay formula
\begin{equation}\label{for:de}
	\begin{aligned}
		\Gamma_i=\gamma_i\alpha k\cdot|c_i|^2,
	\end{aligned}
\end{equation}
where $\gamma_i$ depends on the spatial wave function of final-state products, $\alpha$ describes the coupling strength at decay vertex, and $k$ the momentum of the final state. The coefficient $c_i$ is the probability amplitude for the tetraquark transitioning to the $i$th colorless subsystem, and is quantum mechanically given by the coefficients $c_1$ and $c_8$ when expanding its color-spin wavefunction in terms of the color singlet $1_c$ and color octet $8_c$ components:
\begin{equation}\label{for:ex}
	\begin{aligned}
		|\psi\rangle=c_1|Q_1\bar{q}_2\rangle^{1_c}|Q_3\bar{q}_4\rangle^{1_c}+c_8|Q_1\bar{q}_2\rangle^{8_c}|Q_3 \bar{q}_4\rangle^{8_c}.
	\end{aligned}
\end{equation}

In the heavy quark limit, the values of $\gamma_i$ for the $\Sigma_c$ and the $\Sigma^*_c$ states remains the same approximately \cite{Weng:2019ynv}. For the same reason, one can assume the following approximate relations for the values of $\gamma_i$ connecting the pseudoscalar and vector mesons \cite{Eichten:1979ms, Weng:2021ngd}
\begin{equation}\label{for:gDD}
	\begin{aligned}
		\gamma_{D_sD^*}&=\gamma_{D^*_sD},\gamma_{B_sB^*}=\gamma_{B^*_sB},\gamma_{BD^*}=\gamma_{B^*D},\\
        \gamma_{B_sD^*}&=\gamma_{B^*_sD},\gamma_{BD^*_s}=\gamma_{B^*D_s},\gamma_{B_sD^*_s}=\gamma_{B^*_sD_s}.
	\end{aligned}
\end{equation}

\subsection{Magnetic moments of DH tetraquarks}
\label{sec:mm}

The magnetic moments of DH tetraquarks is one of important electromagnetic properties of hadrons and offer, in general, an useful insight into internal structure of them. In this subsection, we will explore the magnetic moment of a DH tetraquark via computing all contributions from the magnetic moments of constituents in them. At the quark level, the total magnetic moment $\hat{\mu}_{\text{TOT}}$ of a tetraquark is composed of that of total spin $\hat{\mu}_S$ and that of total orbital $\hat{\mu}_L$ of the constituent quarks in hadron. one then has, explicitly

\begin{equation}\label{for:mu}
	\begin{aligned}
		&\hat{\mu}_{\text{TOT}}=\hat{\mu}_{S}+\hat{\mu}_L,\\
		&\hat{\mu}_S=\sum\nolimits_ig_i\mu_i\hat{S}_{i,z},\\	
		&\hat{\mu}_L=\sum\nolimits_{i<j}\left(\frac{m^\prime_i}{m^\prime_i+m^\prime_j}\mu_j+\frac{m^\prime_j}{m^\prime_i+m^\prime_j}\mu_i\right)\hat{L}_{ij,z},
	\end{aligned}
\end{equation}
where the factor $g_i=2$, $\hat{S}_{i,z}$ is the $z$-component of spin of the $i$th quark, and $\hat{L}_{ij,z}$ the $z$-component of the relative orbital angular momentum $\hat{L}_{ij}$ between the $i$th and $j$th quarks. Here, $m^\prime_i\equiv m_i+E_i$ stands for the effective quark mass.

Ignoring the abnormal magnetic moments of quarks, the quark with flavor $f$ will has a magnetic moment of $\mu_f=Q_f/2m^\prime_f$ with $Q_f$ its charge of  constituent quark. The magnetic moment of a DH tetraquark state is then
\begin{equation}\label{for:muN}
	\begin{aligned}
		\mu=\langle\psi|\hat{\mu}_{\text{TOT}}|\psi\rangle,
	\end{aligned}
\end{equation}
where $|\psi\rangle$ stands for the hadron wavefunction consisting of four components
\begin{equation}\label{for:wave}
	\begin{aligned}
        \psi=\left|\phi_\text{color}\otimes\chi_\text{spin}\otimes\eta_\text{flavor}\otimes R_\text{radial}\right\rangle.
	\end{aligned}
\end{equation}
Note here that we are considering the ground-state (S-wave) DH tetraquarks, for which $\hat{L}_{ij}$ is vanishing, the total magnetic moment in Eq. (\ref{for:muN}) equals to the total magnetic moment $\hat{\mu}_S$ of hadron spin, which consists of the spin magnetic moment of the quarks in hadrons. Thus, the radial part of the wavefunction (\ref{for:wave}) will be normalized to unity and the total magnetic moment in Eq. (\ref{for:muN}) depends merely on the color-spin wavefunctions of hadrons, given that its flavor content is specified.

In contrast with the conventional mesons or baryons, DH tetraquarks may have two independent color structures, $6_c \otimes \bar{6}_c$ and $\bar{3}_c \otimes 3_c$, each of them is possible, except for the situation that it is dynamically prohibited. For completeness, one can write the color part of the wavefunction $|\psi\rangle$ of a tetraquark as
\begin{equation}
\phi _{1}=\left\vert {\left( Q_{1}Q_{2}\right) }^{6}{\left( \bar{q}_{3}%
\bar{q}_{4}\right) }^{\bar{6}}\right\rangle ,\quad
\phi _{2}=\left\vert {%
\left( Q_{1}Q_{2}\right) }^{\bar{3}}{\left( \bar{q}_{3}\bar{q}_{4}\right) }%
^{3}\right\rangle ,  \label{colorT}
\end{equation}
and write the spin part of $|\psi\rangle$ as one of the following six states:
\begin{equation}
\begin{aligned}
\chi_{1}^{2,2}={\left| {\left(Q_{1}Q_{2}\right)}_{1}
{\left(\bar{q}_{3}\bar{q}_{4}\right)}_{1} \right\rangle}_{2}, \quad
\chi_{2}^{1,1}={\left| {\left(Q_{1}Q_{2}\right)}_{1}
{\left(\bar{q}_{3}\bar{q}_{4}\right)}_{1} \right\rangle}_{1}, \\
\chi_{3}^{0,0}={\left| {\left(Q_{1}Q_{2}\right)}_{1}
{\left(\bar{q}_{3}\bar{q}_{4}\right)}_{1} \right\rangle}_{0}, \quad
\chi_{4}^{1,1}={\left| {\left(Q_{1}Q_{2}\right)}_{1}
{\left(\bar{q}_{3}\bar{q}_{4}\right)}_{0} \right\rangle}_{1}, \\
\chi_{5}^{1,1}={\left| {\left(Q_{1}Q_{2}\right)}_{0}
{\left(\bar{q}_{3}\bar{q}_{4}\right)}_{1} \right\rangle}_{1}, \quad
\chi_{6}^{0,0}={\left| {\left(Q_{1}Q_{2}\right)}_{0}
{\left(\bar{q}_{3}\bar{q}_{4}\right)}_{0} \right\rangle}_{0},
\end{aligned}
\label{spinT}
\end{equation}
or combination of them, for which we detailed in Appendix \ref{sec:scw}.

Firstly, we consider magnetic moment of the isoscalar $T^+_{cc}$ state with $J^P=1^+$, as a example, for which the color-spin-flavor wavefunction takes the form of
\begin{equation}\label{for:X1}
	\begin{aligned}
        T_{cc}(01^+)=\left[R_1\left|\frac{1}{\sqrt{2}}(\uparrow\uparrow\uparrow\downarrow-\uparrow\uparrow\downarrow\uparrow)\right\rangle|\phi_2\rangle+R_2\left|\frac{1}{\sqrt{2}}(\uparrow\downarrow\uparrow\uparrow-\downarrow\uparrow\uparrow\uparrow)\right\rangle|\phi_1\rangle\right]|cc[\bar{u}\bar{d}]\rangle,
	\end{aligned}
\end{equation}
with $R_1$ the mixing coefficient for the color component $\bar{3}_c \otimes 3_c$ and $R_2$ the mixing coefficient for the component $6_c \otimes \bar{6}_c$. In our framework, these coefficients are determined by the color-spin wavefunction which diagonalize the mass matrix (\ref{for:Ha}) and thereby diagonalize the chromomagnetic interaction matrix (\ref{for:cmi}). Mathematically, two mixing coefficients $R_{1,2}$ forms an eigenvector $(R_1, R_2)^{T}$ which diagonalizes the CMI matrix (\ref{for:cmi}).

Substitution of the wavefunction (\ref{for:wave}) solved as stated above into Eq. (\ref{for:muN}) gives rise to an expression for hadron magnetic moment in terms of the spin magnetic moment of the constituent quarks in hadrons. In the case of the $T^+_{cc}$ with $IJ^{P}=01^{+}$, putting Eq. (\ref{for:X1}) into Eq. (\ref{for:muN}) leads to
\begin{equation}\label{for:T}
	\begin{aligned}
		\mu[T_{cc}(01^+)]&=\left\langle T_{cc}(01^+)\right|\hat{\mu}_{S}\left|T_{cc}(01^+)\right\rangle\\
        &=\frac{R^2_1}{2}\left\langle
        cc[\bar{u}\bar{d}](\uparrow\uparrow\uparrow\downarrow-\uparrow\uparrow\downarrow\uparrow)\right|\hat{\mu}_{S}\left|cc[\bar{u}\bar{d}](\uparrow\uparrow\uparrow\downarrow-\uparrow\uparrow\downarrow\uparrow)\right\rangle\\
        &+\frac{R^2_2}{2}\left\langle
        cc[\bar{u}\bar{d}](\uparrow\downarrow\uparrow\uparrow-\downarrow\uparrow\uparrow\uparrow)\right|\hat{\mu}_{S}\left|cc[\bar{u}\bar{d}](\uparrow\downarrow\uparrow\uparrow-\downarrow\uparrow\uparrow\uparrow)\right\rangle\\
        &=R_1^2\left(\mu_c+\mu_c\right)+R_2^2\left(\mu_{\bar{u}}+\mu_{\bar{d}}\right).
	\end{aligned}
\end{equation}
One sees, evidently, that the magnetic moment of the $T_{cc}$ depends not only on the spin-flavor content but also on its color structures via $R_{1,2}$, which is up to internal chromodynamics of the $T_{cc}$.

Likewise, one can apply similar procedure to the other members of the DH tetraquark $T_{QQ}$ with various color-spin configurations to derive a set of expressions(sum rule) for their magnetic moments, which are the linear combinations of spin magnetic moments of the individual quarks in $T_{QQ}$. The results obtained for the $T_{QQ}$ are
\begin{equation}\label{for:DMD}
	\begin{aligned}
		\mu[\psi^{1,1}_1]&=\frac{\mu_{1234}}{2},\\
		\mu[\psi^{1,1}_2]&=R_1^2\mu_{12^+}+R_2^2\mu_{34^+},\\
		\mu[\psi^{1,1}_3]&=\frac{R_1^2}{2}\mu_{1234}+R_2^2\mu_{12^+}+R_3^2\mu_{34^+}+\sqrt{2}R_1R_2\mu_{34^-},\\
		\mu[\psi^{1,1}_4]&=\frac{R_1^2}{2}\mu_{1234}+R_2^2\mu_{34^+}+R_3^2\mu_{12^+}-\sqrt{2}R_1R_2\mu_{12^-},\\
		\mu[\psi^{1,1}_5]&=R_1^2\mu_{12^+}+\frac{R_2^2}{2}\mu_{1234}+R_3^2\mu_{34^+}-\sqrt{2}R_2R_3\mu_{12^-},\\
		\mu[\psi^{1,1}_6]&=\frac{R_1^2+R_4^2}{2}\mu_{1234}+(R_2^2+R_5^2)\mu_{12^+}+(R_3^2+R_6^2)\mu_{34^+}-\sqrt{2}R_1(R_3\mu_{12^-}-R_2\mu_{34^-})\\
                     &-\sqrt{2}R_4(R_6\mu_{12^-}-R_5\mu_{34^-}),\\
		\mu[\psi^{2,2}_1]&=\mu[\psi^{2,2}_2]=\mu[\psi^{2,2}_3]=\mu_{1234},\\
	\end{aligned}
\end{equation}
where the abbreviations $\mu_{12^\pm}=\mu_{Q_1}\pm\mu_{Q_2}$, $\mu_{34^\pm}=\mu_{\bar{q}_3}\pm\mu_{\bar{q}_4}$, and $\mu_{1234}=\mu_{12^+}+\mu_{34^+}$ are used.

Note that Eqs. (\ref{for:T}) and (\ref{for:DMD}) provide a set of sum rules for the magnetic moments of all DH tetraquark systems, including the nonstrange systems of the $T_{cc}$, the $T_{bb}$ and the $T_{cb}$ . As an usual, we present the values of the magnetic moments in the unit of the nuclear magneton $\mu_N=e/2M_N$, with $M_N=938.272$ MeV the mass of the nucleon \cite{ParticleDataGroup:2024cfk}.

\renewcommand{\tabcolsep}{0.80cm}
\renewcommand{\arraystretch}{1.2}
\begin{table*}[!htbp]
	\caption{The spin-averaged masses and splittings of mesons. All in MeV. Here, $n$ denotes $u/d$ quarks.}
	\label{tab:Spl}
	\begin{threeparttable}
	\begin{tabular}{lclc}
		\bottomrule[1.0pt]\bottomrule[0.5pt]
		Spin-averaged masses											&Values					&Mass splittings					&Values					\\\hline
		$M_{b\bar{b}}=\frac{1}{4}[3M(\Upsilon)+M(\eta_b)]$				&$9445.0$  				&$M(\Upsilon)-M(\eta_b)$			&$61.7$					\\
		$M_{b\bar{c}}=\frac{1}{4}[3M(B^*_c)+M(B_c)]$					&$6316.9\tnote{a}$		&$M(B^*_c)-M(B_c)$					&$56.5\tnote{a}$		\\
		$M_{b\bar{s}}=\frac{1}{4}[3M(B^*_s)+M(B_s)]$					&$5403.3$				&$M(B^*_s)-M(B_s)$					&$48.5$					\\
		$M_{b\bar{n}}=\frac{1}{4}[3M(B^*)+M(B)]$						&$5313.5$				&$M(B^*)-M(B)$						&$45.0$					\\
		$M_{c\bar{c}}=\frac{1}{4}[3M(J/\Psi)+M(\eta_c)]$				&$3068.7$				&$M(J/\Psi)-M(\eta_c)$				&$112.8$				\\
		$M_{c\bar{s}}=\frac{1}{4}[3M(D^*_s)+M(D_s)]$					&$2076.2$				&$M(D^*_s)-M(D_s)$					&$143.9$				\\
		$M_{c\bar{n}}=\frac{1}{4}[3M(D^*)+M(D)]$						&$1971.4$				&$M(D^*)-M(D)$						&$142.0$				\\
		$M_{s\bar{s}}=\frac{1}{4}[3M(\phi)+M(\eta_s)]$					&$936.3\tnote{b}$		&$M(\phi)-M(\eta_s)$				&$332.5\tnote{b}$		\\
		$M_{s\bar{n}}=\frac{1}{4}[3M(K^*)+M(K)]$						&$796.1$				&$M(K^*)-M(K)$						&$397.9$				\\
		$M_{n\bar{n}}=\frac{1}{4}[3M(\rho)+M(\pi)]$						&$615.2$				&$M(\rho)-M(\pi)$					&$640.3$				\\
		\bottomrule[0.5pt]\bottomrule[1.0pt]
	\end{tabular}
	\begin{tablenotes}
		\footnotesize
		\item[a] A mass value $M(B^{*-}_c)=6331(4)(6)$ MeV from the lattice \cite{Mathur:2018epb} is used here for the $B^{*-}_c$ meson.
		\item[b] We have excluded the $\text{SU}(3)_f$ broken effects and introduced a pure $s\bar{s}$ state with mass $M(\eta_s)=686.92$ MeV, determined via
        $M^2(\eta_s)=2M^2_K-M^2_\pi$
        \cite{Witten:1983ut,Feldmann:1998vh,Feldmann:1998su,Scadron:1982eg,Weng:2018mmf}.
	\end{tablenotes}
	\end{threeparttable}
\end{table*}

\subsection{Parameters}
\label{sec:dop}

This subsection is to determine the model parameters in Eq. (\ref{for:Ha}). For this purpose, we maintain the setup of the quark mass, $m_n=230$ MeV, $m_s=328$ MeV, $m_c=1440$ MeV, and $m_b=4480$ MeV, which are extracted from the Regge-like spectrum of the established heavy mesons and baryons \cite{Jia:2019bkr}. For the chromomagnetic couplings $A_{ij}$, we shall resort to hyperfine mass splitting of the conventional mesons or baryons. For the CMI coupling $A_{i\bar{j}}$ of the quark-antiquark pair $i\bar{j}$, we apply the mass formula (\ref{for:Ha}) to the vector ($\mathbb{V}$) and pseudoscalar ($\mathbb{P}$) mesons to obtain, using the notation $V_{ij}\equiv A_{ij}/m_im_j$,
\begin{equation} \label{V-P}
	\begin{aligned}	M(\mathbb{V})&=m^\prime_i + m^\prime_j+B_{i\bar{j}}+\frac{16}{3}V_{i\bar{j}},\\
		M(\mathbb{P})&=m^\prime_i+m^\prime_j+B_{i\bar{j}}-16V_{i\bar{j}},
	\end{aligned}
\end{equation}
where $m^\prime_i=m_i+E_i$ as before. From two equations in Eq. (\ref{V-P}), one obtains the mass splitting between vector and pseudoscalar mesons
\begin{equation}\label{VP}
	\begin{aligned}
		M(\mathbb{V})-M(\mathbb{P})=\frac{64}{3}\frac{A_{i\bar{j}}}{m_im_j},
	\end{aligned}
\end{equation}
from which the CMI couplings $A_{i\bar{j}}$ are extractable. For the light pseudoscalar mesons, for instance, Eq. (\ref{VP}) becomes,
\begin{equation}
	\begin{aligned}
		M(\rho)-M(\pi)=\frac{64}{3}\frac{A_{n\bar{n}}}{(230\ \text{MeV})^2},
	\end{aligned}
\end{equation}
where the LHS is $640.3$ MeV and the up/down quark mass $m_n$ is $230$ MeV. This yields $A_{n\bar{n}}=0.00159\ \text{GeV}^3$ for the up and down quarks. Likewise, one can apply Eq. (\ref{VP}) to other mesons, shown in Table \ref{tab:Spl}, and to extract the corresponding CMI couplings with the help of the measured mass splittings in Table \ref{tab:Spl}. The results are listed in Table \ref{tab:aij}, from which the couplings $A_{i\bar{j}}$ for the quark-antiquark pair $i\bar{j}$ can be evaluated. The obtained results for $A_{i\bar{j}}$ are listed in Table \ref{tab:aij}.

As for the quark-quark CMI couplings $A_{ij}$, one can apply Eq. (\ref{for:Ha}) to the observed baryons with spin-parities $J^{P}=1/2^{+}$ and $J^{P}=3/2^{+}$ to obtain the mass splitting between the octets and decuplets, namely, the analogy of Eq. (\ref{VP}). As a result, all $A_{ij}$ are determined. In the case of the $\Delta$ and $N$ states, for instance, one can use the following mass splitting,
\begin{equation}\label{}
	\begin{aligned}
		M(\Delta)-M(N)=16\frac{A_{nn}}{(230\ \text{MeV})^2},
	\end{aligned}
\end{equation}
where the LHS is $291.7$ MeV. It follows that $A_{nn}=0.00096\ \text{GeV}^3$.

\begin{figure}[htpb]
\begin{center}
\vspace*{0em}
\hspace*{0em}
\includegraphics[scale=0.38]{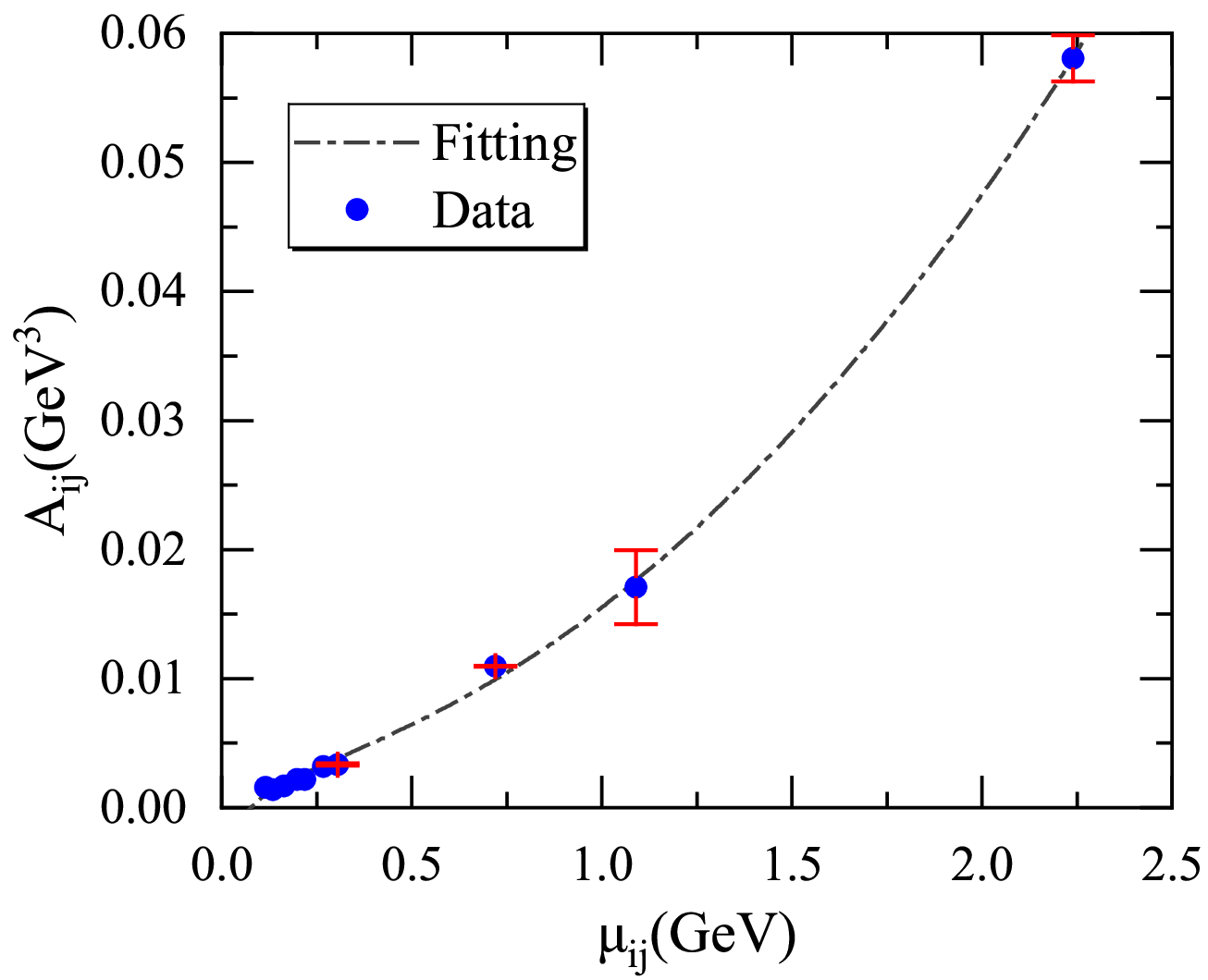}
\caption{The behavior of chromomagnetic coupling (blue point) as a function of reduced mass.}\label{CMICP}
\end{center}
\end{figure}

In the cases where experimental data are lacking($A_{cc}$, for instance), we employ an empirical scaling relation $A_{i\bar{j}}/A_{ij}=3/2\pm0.3$, in Refs. \cite{Weng:2018mmf, Liu:2019zoy}, to evaluate directly them from the known quark-antiquark couplings $A_{i\bar{j}}$. The results are obtained for all CMI couplings and listed in Table \ref{tab:aij}. By the way, we find a simple function to well fit the obtained CMI couplings $A_{i\bar{j}}$, which depends upon the reduced mass $\mu_{ij}$ of the quark pair $i\bar{j}$. This fit-function takes the form
\begin{equation}
    \begin{aligned}
        A_{i\bar{j}}(\mu_{i\bar{j}})=a+b\frac{\mu^2_{i\bar{j}}}{\Lambda^2}+c\text{ln}\left(\frac{\mu^2_{i\bar{j}}}{\Lambda^2}\right),
    \end{aligned}
\end{equation}
with $\Lambda=0.2$ GeV, $a=0.00197$, $b=0.00041$ and $c=0.00104$, shown in FIG. \ref{CMICP}.

\renewcommand{\tabcolsep}{0.20cm}
\renewcommand{\arraystretch}{1.2}
\begin{table*}[!htbp]
	\caption{The chromomagnetic couplings $A_{ij}$ between any two quarks. All in $\text{GeV}^3$.}
	\label{tab:aij}
	\begin{tabular*}{\linewidth}{cccccccccc}
		\bottomrule[1.0pt]\bottomrule[0.5pt]
		$A_{n\bar{n}}$		&$A_{s\bar{n}}$		&$A_{s\bar{s}}$		&$A_{c\bar{n}}$		&$A_{c\bar{s}}$	&$A_{c\bar{c}}$	&$A_{b\bar{n}}$	&$A_{b\bar{s}}$	&$A_{b\bar{c}}$	&$A_{b\bar{b}}$	\\
		$0.00159$			&$0.00141$			&$0.00168$			&$0.0022$			&$0.00318$		&$0.01096$		&$0.00217$		&$0.00334$		&$0.01709$		&$0.05805$		\\
		$A_{nn}$			&$A_{ns}$			&$A_{ss}$			&$A_{nc}$			&$A_{sc}$		&$A_{cc}$		&$A_{nb}$		&$A_{sb}$		&$A_{cb}$		&$A_{bb}$		\\
		$0.00096$			&$0.00094$			&$0.00112$			&$0.00147$			&$0.00212$		&$0.00731$		&$0.00145$		&$0.00223$		&$0.0114$		&$0.0387$		\\
		\bottomrule[0.5pt]\bottomrule[1.0pt]
	\end{tabular*}
\end{table*}

To evaluate the interquark binding energies $B_{Q\bar{Q}^\prime}$, which is relevant (nonvanishing) for the hadrons containing two or more heavy(strange) quarks, we use the relations proposed in Ref. \cite{Liu:2023vrk}, which is to extract the binding energies $B_{Q\bar{Q}^\prime}$, stemming mainly from the short-distance color interaction, via concealing the strings connecting quarks $Q\bar{Q}^\prime$, responsible for long-distance color interaction, and effective quark masses in hadrons based on the QCD picture of the hadrons. For the quark pair $Q\bar{Q}^\prime$ in the color singlet $1_c$, they take the form of,
\begin{equation}\label{for:BQQ}
	\begin{aligned}
        B_{c\bar{s}}&=M_{c\bar{s}}-M_{c\bar{n}}-M_{s\bar{n}}+M_{n\bar{n}}=-76.0\text{ MeV},\\
        B_{b\bar{s}}&=M_{b\bar{s}}-M_{b\bar{n}}-M_{s\bar{n}}+M_{n\bar{n}}=-91.1\text{ MeV},\\
		B_{c\bar{c}}&=M_{c\bar{c}}-2M_{c\bar{n}}+M_{n\bar{n}}=-258.8\text{ MeV},\\
		B_{b\bar{b}}&=M_{b\bar{b}}-2M_{b\bar{n}}+M_{n\bar{n}}=-566.8\text{ MeV},\\
		B_{c\bar{b}}&=M_{b\bar{c}}-M_{b\bar{n}}-M_{c\bar{n}}+M_{n\bar{n}}=-352.8\text{ MeV}.
	\end{aligned}
\end{equation}
Here, the energy of string connecting two heavy quarks is assumed to be negligible and the energy of string connecting the heavy quark and light antiquarks $Q\bar{n}^\prime$ is half of that of string connecting the nonstrange quark and antiquarks $n\bar{n}^\prime$ (due to the difference of the local position of the center-of-mass of hadron between two situations).

For the quark-quark pair $QQ^\prime$ in other color configurations, we expect $B_{QQ^\prime}$ to be governed by the short-distance interquark force, which is of the Coulomb-like according to the perturbative QCD, and thereby it scales like the color-factor $\langle\bm{\lambda}_Q\cdot\bm{\lambda}_{Q^\prime}\rangle$ given in Eq. (\ref{factors}). We apply then the scaling of the color-factor in Ref. \cite{Zhang:2021yul} to estimate the binding energies for heavy quark pair $QQ^\prime$ based on the binding $B_{Q\bar{Q}^\prime}$ for $QQ^\prime$ in color singlet in Eq. (\ref{for:BQQ}). In the case of the heavy pair $QQ^\prime$ in the color antitriplet $\bar{3}_c$, for instance, the scaling factor $g(\bar{3}_c)$ is 1/2, which leads to $B_{\bar{3}_c}/B_{1_c}=1/2$ for given $QQ^\prime$. One can then estimate the quark-quark binding energies in $\bar{3}_c$ to be (the half of the respective values in Eq. (\ref{for:BQQ}) )
\begin{equation}\label{for:Bdi}
	\begin{aligned}
		&B_{cc}=-129.4\ \text{MeV},B_{bb}=-283.4\ \text{MeV},B_{cb}=-176.4\ \text{MeV}.
	\end{aligned}
\end{equation}
For the other color rep. $R$ of the quark pair $Q\bar{Q}^\prime$ or $QQ^\prime$, one can first compute the scaling factor \cite{Zhang:2021yul}
 \begin{equation}
	\begin{aligned}
        g(R)\equiv \frac{\langle\bm{\lambda}_Q\cdot\bm{\lambda}_{Q(\bar{Q}^\prime)}\rangle_{R}}{\langle\bm{\lambda}_{Q}\cdot\bm{\lambda}_{\bar{Q}^\prime}\rangle_{1_c}}
        =\frac{(B_{QQ^\prime(\bar{Q}^\prime)})_R}{(B_{Q\bar{Q}^\prime})_{1_c}},
	\end{aligned}
\end{equation} 
and then use it to scale $(B_{Q\bar{Q}^\prime})_{1_c}$ to either $(B_{QQ^\prime})_R$ or $(B_{Q\bar{Q}^\prime})_R$.

The third parameter to be determined is the string energy $E_{i}$, which stems from the long-distance interaction and should be of flavor independent in principle.
In hadrons with finite size, however, the string lives between the quark sources and is constrained by hadron size due to the kinematics of the sources(with finite masses). Thus, string energy $E_{i}$ displays a slight dependence on flavor of the quarks $i$ when averaging in hadron. In $T_{QQ}$, we assume the string energy $E_{n}$ is same for the $u$ and $d$ quarks, while the string energies $E_{s}$, $E_{c}$ and $E_{b}$ differ from $E_n$ and differ from each other as the $s$, $c$ and $b$ quarks, one heavier than the other. In the following, we fix these parameters via applying Eq. (\ref{for:Ha}) to the relevant hadrons.
\begin{itemize}
	\item [i)] $E_n$: Experimentally, the spin-averaged mass of the nucleons $\Delta$ and $N$ is
	\begin{equation}\label{for:De}
		\begin{aligned}
			\frac{4}{6}M(\Delta)+\frac{2}{6}M(N)=1134.1\ \text{MeV}.
		\end{aligned}
	\end{equation}
Application of Eq. (\ref{for:Ha}) to the $\Delta$ and $N$ composed of the $nnn$ system with spins $3/2$ and $1/2$ leads to
    \begin{equation}
		\begin{aligned}
			M(\Delta)=3m^\prime_n+8V_{nn},M(N)=3m^\prime_n-8V_{nn},
		\end{aligned}
	\end{equation}
respectively. Putting the above relations into Eq. (\ref{for:De}) and solving for $E_n$, one obtains $E_n=131.8$ MeV.
	\item [ii)] $E_s$: Averaging the masses of the strange baryons $\Sigma^*$ and $\Sigma$ gives, experimentally,
	\begin{equation}\label{for:Ds}
	   \begin{aligned}
	       \frac{4}{6}M(\Sigma^*)+\frac{2}{6}M(\Sigma)=1320.0\text{ MeV}.
	   \end{aligned}
	\end{equation}	
Application of Eq. (\ref{for:Ha}) to the strange baryons $\Sigma^*$ and $\Sigma$ of the $nns$ system with spins $3/2$ and $1/2$ gives rise to
	\begin{equation}\label{for:Ds1}
		\begin{aligned}
			M(\Sigma^*)&=2m^\prime_n+m^\prime_s+(8/3)(V_{nn}+2V_{ns}),\\
			M(\Sigma)&=2m^\prime_n+m^\prime_s+(8/3)(V_{nn}-4V_{ns}),\\
		\end{aligned}
	\end{equation}
Solving Eqs. (\ref{for:Ds}) combined with Eq. (\ref{for:Ds1}) for $E_s$, one gets $E_s=219.8$ MeV.
	
	\item [iii)] $E_c, E_b$: In average, the observed charmed baryon $\Sigma^{(*)}_c=nnc$ and the bottom baryon $\Sigma^{(*)}_b=nnb$ have the masses, \begin{equation}\label{for:Dc}
		\begin{aligned}
			\frac{4}{6}M(\Sigma^*_c)+\frac{2}{6}M(\Sigma_c)&=2496.9\text{ MeV},\\
		\end{aligned}
	\end{equation}
	\begin{equation}\label{for:Db}
		\begin{aligned}
			\frac{4}{6}M(\Sigma^*_b)+\frac{2}{6}M(\Sigma_b)&=5826.1\text{ MeV},\\
		\end{aligned}
	\end{equation}
respectively. Similar precedure applying Eq. (\ref{for:Ha}) to them gives $E_c=284.7$ MeV and $E_b=573.8$ MeV, respectively.
\end{itemize}

\section{Doubly heavy tetraquarks}
\label{sec:CD}

Given the parameters determined, we apply our approach in Sect. \ref{sec:for} to the nonstrange and strange DH tetraquarks of the four-quark system $QQ^\prime\bar{q}\bar{q}$, which consist of two heavy quarks ($Q,Q^\prime=c, b$) and two light antiquarks $\bar{q}\bar{q}$ with $q=u, d, s$. For abbreviation, we use notation $T_{QQ^\prime}(M,IJ^P)$ to denote DH tetraquark with mass $M$ and  the quantum number of the isospin and spin-parity $IJ^P$.

We firstly solve the CMI model (\ref{for:Ha}) for nonstrange DH tetraquarks $QQ^\prime \bar{n}\bar{n}$ and then for strange DH tetraquarks $QQ^\prime \bar{s}\bar{n}$ and $QQ^\prime \bar{s}\bar{s}$. As stated in Sect. \ref{sec:for}, for this, one has to use the proper wavefunction (\ref{for:wave}) constructed in the color-spin-flavor space, as illustrated in Eq. (\ref{for:X1}), to diagonalize the mass Hamiltonian, that is, to diagonalize the $B+H_{\text{CMI}}$ with $H_{\text{CMI}}$ given in Eq. (\ref{for:cmi}). Next, one can substitute the obtained eigenvalues of the Hamiltonian $B+H_{\text{CMI}}$ into Eq. (\ref{for:Ha}) to obtain the 1S-wave masses of the relevant DH tetraquark systems with given quantum numbers. The obtained eigenvectors provide the respective mixing coefficients $R_{i}$ of the given configurations and can be used to compute via Eq. (\ref{for:DMD}) the the magnetic moments of the hadron systems.

When the DH tetraquarks are above the two-meson thresholds, one can employ the two-body formula (\ref{for:de}) to explore the relative strong decay widths of them. In addition, When the DH tetraquarks, some of the doubly-bottom and bottom-charmed tetraquarks in our case, are below the two-meson thresholds, they decay mainly via the weak decay or radiative decay. Thus, lifetime of these doubly-bottom and bottom-charmed tetraquarks is due to the shorter one of the lifetimes by the weak decay and the radiative decay. The lifetime can stem from the radiative decay when it is easier averagely to occur to end the life of the DH tetraquarks before the weak decay.

For the weak decays, a crude estimate of lifetime is available using the procedure in Ref. \cite{Bjorken:1985ei,Moinester:1995fk,Fleck:1989mb,Gronau:2010if,Karliner:2014gca} for the weak transition involving the most-favored Cabibbo-Kobayashi-Maskawa matrix elements (at $b \rightarrow cW^{*-}$ )
\begin{equation}\label{for:Gan}
	\begin{aligned}
        \Gamma(T_{bb})=\frac{18G_F^2M_{i}^5}{192\pi^3}F(x)|V_{cb}|^2,
	\end{aligned}
\end{equation}
which depends on the squared mass ratio $x=[M_{f}/M_{i}]^{2}$ of the final state and initial state via a kinematic suppression factor \cite{Quang:1998yw,Gronau:2010if}
\begin{equation}\label{for:ksf}
	\begin{aligned}
		F(x)=1-8x+8x^3-x^4+12x^2\text{ln}(1/x),
	\end{aligned}
\end{equation}
with $|V_{cb}|=0.042$ the Cabibbo-Kobayashi-Maskawa matrix element \cite{ParticleDataGroup:2024cfk} and $G_F=1.166\times10^{-5}\ \text{GeV}^{-2}$ \cite{ParticleDataGroup:2024cfk} the Fermi constant.

When the radiative decay dominates for the states considered, one can use the M1 radiative decay formula \cite{Wang:2023bek},
\begin{equation}\label{for:Ga2}
	\begin{aligned}
        &\Gamma=\frac{\omega^3}{M^2_{p}}\frac{\alpha_\text{EM}}{2J_{i}+1}\frac{|\mu_{T_{i}\to T_{f}}|^2}{\mu^2_N},\\
        &J_z=\text{Min}\{J_{i},J_{f}\},\\
        &\mu_{T_{i}\to T_{f}}=\left\langle J_{f},J_z|\hat{\mu}_{\text{TOT}}|J_{i},J_z\right\rangle,\\
	\end{aligned}
\end{equation}
where $\alpha_\text{EM}$ is the fine-structure constant, $\mu_{T_{i}\to T_{f}}$ is the electromagnetic transition magnitude, $J_{i}$ is the total angular momentum of the initial state $T_{i}$ and $J_{f}$ is that of the final state $T_{f}$, $\omega=\frac{M^2_{i}-M^2_{f}}{2M_{i}}$ the momentum magnitude of the outgoing photon during the radiative decay.

\subsection{The doubly charmed states}
\label{sec:cc}

\textit{Nonstrange tetraquarks} Consider first the DH tetraquark $T_{cc}=cc\bar{n}\bar{n}$, composed of two charm quarks $cc$ and two light nonstrange antiquarks $\bar{n}\bar{n}$. As stated at beginning of Sect. \ref{sec:CD}, we utilize the wavefunctions (\ref{for:X1}) and others detailed in Appendix \ref{sec:scw} for the DH tetraquark without strange quarks to diagonalize the Hamiltonian $B+H_{\text{CMI}}$ and put the obtained eigenvalues of $B+H_{\text{CMI}}$ into Eq. (\ref{for:Ha}) to give the masses of the relevant tetraquark systems. The numerical results for the CMI matrices and the whole Hamiltonian are given in Table \ref{tab:c1}, and the masses and magnetic moments obtained thereby are listed in Table \ref{tab:2} for the various states of the nonstrange $T_{cc}$ systems, including the cases of color-spin states mixed by chromomagnetic interaction. We compute similar results for all other $T_{cc}$ with strangeness $S=0,-1,-2$ and list them in Tables \ref{tab:2}, \ref{tab:3}, and \ref{tab:4}, respectively.

\renewcommand{\tabcolsep}{0.55cm}
\renewcommand{\arraystretch}{1.2}
\begin{table*}[!htbp]
	\caption{The computed matrices of the CMI and the whole Hamiltonian of the doubly charmed tetraquarks $cc\bar{n}\bar{n}$, $cc\bar{s}\bar{n}$, and $cc\bar{s}\bar{s}$.}
\label{tab:c1}
\begin{tabular}{crcc}
	\bottomrule[1.0pt]\bottomrule[0.5pt]
Systems       &$I(J^P)$   	& CMI matrices	& Hamiltonians	\\\hline
		$cc\bar{n}\bar{n}$  &$0(1^+)$
&$\begin{bmatrix}-136.5&-75.3\\-75.3&-10.2\\\end{bmatrix}$
&$\begin{bmatrix}-265.9&-75.3\\-75.3&54.5\\\end{bmatrix}$\\
                    		&$1(0^+)$
&$\begin{bmatrix}-13.0&130.4\\130.4&87.0\\\end{bmatrix}$
&$\begin{bmatrix}-142.4&130.4\\130.4&151.7\\\end{bmatrix}$\\
                    		&$1(1^+)$       	&$22.5$	&$-106.9$				\\
                    		&$1(2^+)$        	&$93.5$	&$-35.9$				\\
        $cc\bar{s}\bar{n}$  &$1/2(0^+)$
&$\begin{bmatrix}-28.9&131.3\\131.3&63.8\\\end{bmatrix}$
&$\begin{bmatrix}-234.3&131.3\\131.3&-61.4\\\end{bmatrix}$\\
                    		&$1/2(1^+)$
&$\begin{bmatrix}6.8&0.3&0.7\\0.3&-90.1&-75.8\\0.7&-75.8&-2.5\\\end{bmatrix}$
&$\begin{bmatrix}-198.6&0.3&0.7\\0.3&-295.5&-75.8\\0.7&-75.8&-127.7\\\end{bmatrix}$\\
                    		&$1/2(2^+)$        	&$78.3$	&$-127.1$				\\
		$cc\bar{s}\bar{s}$  &$0(0^+)$
&$\begin{bmatrix}-34.8&132.1\\132.1&55.7\\\end{bmatrix}$
&$\begin{bmatrix}-316.2&132.1\\132.1&-259.6\\\end{bmatrix}$\\
                    		&$0(1^+)$       	&$1.1$	&$-280.2$				\\
                    		&$0(2^+)$        	&$73.1$	&$-208.3$				\\
		\bottomrule[0.5pt]\bottomrule[1.0pt]
	\end{tabular}
\end{table*}

To illustrate the computation, we choose the $T_{cc}$ state with $I(J^P)=0(1^+)$ as an example. Using the parameters determined in Sect. \ref{sec:for}, one obtains numerically for the matrix $H$ and $H_{\text{CMI}}$,
\begin{equation}\label{for:HB}
	\begin{aligned}
		H=B+H_{\text{CMI}}=\begin{bmatrix}-265.9&-75.3\\-75.3&54.5\\\end{bmatrix};\ H_{\text{CMI}}=\begin{bmatrix}-136.5&-75.3\\-75.3&-10.2\\\end{bmatrix},
	\end{aligned}
\end{equation}
from the later of which one gets two eigenvalues $h_{1,2}$ and the respective eigenvectors $R_{1,2}$ for $H$,
\begin{equation}\label{for:Eig}
	\begin{aligned}
		h_{1}=-282.7, R_{1}=\begin{bmatrix} -0.976\\-0.218\\  \end{bmatrix} ; \ h_{2}=71.3, R_{2}=\begin{bmatrix} 0.218\\-0.976\\  \end{bmatrix}  .
	\end{aligned}
\end{equation}
With $2(m^\prime_c+m^\prime_n)=2(1440+284.7)+2(230+131.8)= 4173.0$ MeV, substitution of two eigenvalues ($h_{1,2}$) into Eq. (\ref{for:Ha}), namely, $M=2(m^\prime_c+m^\prime_n)+\text{Eig}(H)$, gives rise to two masses for two color-spin multiplets $T_{cc}$ and $T^\prime_{cc}$:
\begin{equation}\label{for:M}
	\begin{aligned}
		&M(T_{cc},01^+)=4173.0\ -282.7\ =3890.3\ \text{MeV},\\
		&M(T^\prime_{cc},01^+)=4173.0\ +71.3=4244.3\ \text{MeV}.
	\end{aligned}
\end{equation}

The lower color-spin state $T_{cc}(01^+)$ is near the $DD^*$ mass threshold $M(DD^*)=3875.8$ MeV, from the above with the spacing about $15$ MeV. With Eq. (\ref{for:X1}), its wavefunction takes the form of
 \begin{equation}\label{for:X}
	\begin{aligned}
        T_{cc}(01^+)=\left[-0.976\left|\frac{1}{\sqrt{2}}(\uparrow\uparrow\uparrow\downarrow-\uparrow\uparrow\downarrow\uparrow)\right\rangle|\phi_2\rangle-0.218\left|\frac{1}{\sqrt{2}}(\uparrow\downarrow\uparrow\uparrow-\downarrow\uparrow\uparrow\uparrow)\right\rangle|\phi_1\rangle\right]|cc[\bar{u}\bar{d}]\rangle,
	\end{aligned}
\end{equation}
in which $R_1=-0.976$ is the amplitude of the color configuration $\bar{3}_c \otimes 3_c$ and $R_2=-0.218$ the amplitude for the $6_c \otimes \bar{6}_c$ mixed in $T_{cc}(01^+)$. This gives mixing weight $(R_1)^{2}=0.95$ for the exotic component $6_c \otimes \bar{6}_c$ mixed in $T_{cc}(01^+)$.

\renewcommand{\tabcolsep}{0.55cm}
\renewcommand{\arraystretch}{1.2}
\begin{table*}[!htbp]
	\caption{Masses (in MeV), mixing weights and magnetic moments (for $I=0$ or $I_{z}=1,0,-1$, in $\mu_N$) of non-strange $T_{cc}$ states. $dM$ is mass spacing relative to the relevant two-meson threshold. The column 5 shows weights of the configuration $\bar{3}_c\otimes3_c$ mixed in hadrons.}
	\label{tab:2}
	\begin{tabular}{cclccc}
		\bottomrule[1.0pt]\bottomrule[0.5pt]
		$I$      		&$J^P$     		&Masses	(MeV)	&$dM$			&Mixings ($\%$)	&Magnetic moments	\\\hline
		$0$      		&$1^+$      	&$3890\pm13$	&$14$				&$95$			&$0.65$				\\
						&				&$4244\pm15$	&$368$				&$5$			&$-0.79$			\\
		$1$     		&$0^+$       	&$3981\pm14$	&$246$				&$87$			&					\\
						&				&$4374\pm14$	&$636$				&$13$			&					\\
						&$1^+$       	&$4066\pm13$	&$190$				&$100$			&$1.23,-0.07,-1.37$	\\
						&$2^+$      	&$4137\pm13$	&$120$				&$100$			&$2.45,-0.14,-2.73$	\\
		\bottomrule[0.5pt]\bottomrule[1.0pt]
	\end{tabular}
\end{table*}

Considering the uncertainty of about 13 MeV in our parameters, our mass prediction in Eq. (\ref{for:M}) is not precise enough to exclude possibility that it is a narrow resonance structure. This is similar to the situation of the tetraquark $T_{cc}(3875)^+$ having mass and width \cite{LHCb:2021vvq, LHCb:2021auc}
\begin{equation}
	\begin{aligned}
		M=3874.8\pm0.1\ \text{MeV}\:\Gamma=0.4\pm0.2\ \text{MeV},
	\end{aligned}
\end{equation}
reported by LHCb. Given $R_{1,2}=-0.976(0.218)$, one has for the magnetic moments,
\begin{equation}\label{for:mT}
	\begin{aligned}
		\mu\left[T^+_{cc},01^+\right]&=0.65\ \mu_N,\\
        \mu\left[T^{+\prime}_{cc},01^+\right]&=-0.79\ \mu_N,
	\end{aligned}
\end{equation}
which differs in chromodynamics: the former contributes $95\%$ from two charm quarks, while the later of $95\%$ from the $\bar{u}$ and $\bar{d}$ antiquarks. Our prediction of the magnetic moments for the tetraquark $T^+_{cc}$ with $I(J^P)=0(1^+)$ are compared to other computations cited in the compact and molecule pictures (see Table \ref{tab:3875}).

Likewise, the matrix $H_{\text{CMI}}$ and $H$ for other higher nonstrange $T_{cc}$ states with $I(J^{P})=1(0,1,2)^{+}$ can be computed similarly, with results for $H_{\text{CMI}}$ and $H$ shown in Table \ref{tab:c1}. As such, the masses and mixing weights can be solved numerically and are listed explicitly in Table \ref{tab:2}. The magnetic moments for other higher nonstrange $T_{cc}$ states are listed in Table \ref{tab:2}. Evidently, the magnetic moment depends crucially on the color structure of the tetraquarks, being helpful for revealing the internal structure of hadrons.

\textit{Strange tetraquarks} For the strange tetraquark $T_{cc,s}$ and $T_{cc,ss}$, the masses and mixing weights are computed and listed in Table \ref{tab:3} and \ref{tab:4}, based on the mass matrices in Table \ref{tab:c1}. The magnetic moments for them are shown in Table \ref{tab:3} and \ref{tab:4}. Evidently, most of the strange tetraquarks are well above the relevant two-meson thresholds. Among them, the lowest state, $T_{cc,s}(4034,(1/2)1^+)$, is $55\pm31$ MeV above the $(D_sD)^*$ mass threshold.
For its strong decay channels $D^*_sD$ and $D_sD^*$, one can use Eq. (\ref{for:de}) combined with Eq. (\ref{for:gDD}) to obtain the width ratio between two channels,
\begin{equation}\label{for:4034}
	\begin{aligned}
		\frac{\Gamma(T_{cc,s}(4034,(1/2)1^+)\to D_sD^*)}{\Gamma(T_{cc,s}(4034,(1/2)1^+)\to D_s^*D)}=\frac{59.21\gamma_{D_sD^*}\alpha}{56.68\gamma_{D_s^*D}\alpha}=1.04,
	\end{aligned}
\end{equation}
and find them to be close each other in strong interaction.

Note that the meson $D^*$ has strong tendency, with largest branching fraction $Br=(67.7\pm0.5)\%$ \cite{ParticleDataGroup:2024cfk}, to decay to the final state $D\pi$. This allows the strong decay process $T_{cc,s}\to D_s(D\pi)^{J=1}$ in P-wave. However, this channel is suppressed by the kinematic factor, and thus is not the dominant decay channel for the strange $T_{cc,s}$. We suggest that the strange tetraquark $T_{cc,s}(4034,(1/2)1^+)$ is more likely to be find in the channels of $T_{cc,s}\to D_sD^*$ and $D_s^*D$, and this state has two magnetic moments (in $\mu_N$), $\mu[T^{++}_{cc,s},(1/2)1^+]=0.82$ and $\mu[T^+_{cc,s},(1/2)1^+]=0.50$, with respect to two isospin components $I_{z}=1/2,-1/2$, respectively.

\renewcommand{\tabcolsep}{0.52cm}
\renewcommand{\arraystretch}{1.2}
\begin{table*}[!htbp]
	\caption{Masses (in MeV), mixing weights of the rep. $\bar{3}_c\otimes3_c$ and magnetic moments (for $I_{z}=1/2,-1/2$, in $\mu_N$) of the tetraquark $T_{cc,s}$ with strangeness $S=-1$. $dM$ is the mass spacing relative to the two-meson threshold. $M(DD^*_s)=3979$ MeV.}
	\label{tab:3}
	\begin{tabular}{cclccc}
		\bottomrule[1.0pt]\bottomrule[0.5pt]
		$I$   		         	&$J^P$			&Masses	(MeV)	&$dM$	&Mixings ($\%$)&Magnetic moments ($I_{z}$)	\\\hline
		$1/2$ 			   		&$0^+$  		&$4054\pm23$	&$218$		&$77$		&					\\
								&				&$4368\pm26$	&$532$		&$23$		&					\\
								&$1^+$  		&$4034\pm31$	&$55$		&$87$		&$0.82,0.50$		\\
								&				&$4160\pm4$	    &$181$		&$100$		&$1.08,-0.24$		\\
								&				&$4260\pm14$	&$281$		&$13$		&$1.34,-0.91$		\\
								&$2^+$  		&$4232\pm22$	&$111$		&$100$		&$2.16,-0.43$		\\
		
		\bottomrule[0.5pt]\bottomrule[1.0pt]
	\end{tabular}
\end{table*}

\renewcommand{\tabcolsep}{0.55cm}
\renewcommand{\arraystretch}{1.2}
\begin{table*}[!htbp]
	\caption{Masses (in MeV), mixing weights of the rep. $\bar{3}_c\otimes3_c$ and magnetic moments $\mu$ (for $I=0$, in $\mu_N$) of the tetraquark $T_{cc,ss}$ with strangeness $S=-2$. $dM$ is the mass spacing relative to the two-meson threshold.}
	\label{tab:4}
	\begin{tabular}{cclccc}
		\bottomrule[1.0pt]\bottomrule[0.5pt]
		$I$            	&$J^P$			&Masses (MeV)	&$dM$	    &Mixings ($\%$)	&Magnetic moments ($I_{z}$)		\\\hline
		$0$  	  		&$0^+$  		&$4122\pm22$	&$185$		&$60$			&						\\
						&				&$4392\pm23$	&$455$		&$40$			&						\\
						&$1^+$			&$4265\pm20$	&$184$		&$100$			&$0.93$					\\
						&$2^+$			&$4337\pm20$	&$113$		&$100$			&$1.87$					\\
		\bottomrule[0.5pt]\bottomrule[1.0pt]
	\end{tabular}
\end{table*}

\renewcommand{\tabcolsep}{0.12cm}
\renewcommand{\arraystretch}{1.2}
\begin{table}
	\caption{Comparison of predictions for the magnetic moments (in $\mu_N$) of the $T^+_{cc}(01^+)$ in the compact and molecular pictures.}
	\label{tab:3875}
	\begin{tabular}{lcc}
		\bottomrule[1.0pt]\bottomrule[0.5pt]
		References                          	& Magnetic moments						& Methods    \\ \hline
		\multicolumn{3}{l}{Compact}\\
		Present work			                & $0.65$                                &\\
		\cite{Zhang:2021yul}              		& $0.88$              					&MIT bag model\\
		\cite{Deng:2021gnb}                     & $0.18$                  				&Color screening confinement\\
		&										&+ One-Gluon-Exchange \\
		\cite{Azizi:2021aib}                   	& $0.66^{+0.34}_{-0.23}$                &QCD sum rule\\
		\cite{Wu:2022gie}						& $0.73$								&Constituent quark model\\
		\cite{Mutuk:2023oyz}					& $0.28$								&Diffusion Monte Carlo\\
		\multicolumn{3}{l}{Molecular}\\
		\cite{Deng:2021gnb}                     & $0.13$                  				&One-Boson-Exchange\\
		\cite{Lei:2023ttd}						& $-0.09$								&Constituent quark model\\
		\cite{Azizi:2021aib}					& $0.43^{+0.23}_{-0.22}$				&QCD sum rule\\
		\bottomrule[0.5pt]\bottomrule[1.0pt]
	\end{tabular}
\end{table}

\subsection{The doubly bottom states}
\label{sec:bb}

We consider the doubly bottom tetraquark state, denoted as $T_{bb}(M,IJ^P)$ when it has the mass $M$ and isospin $I$ and spin-parity $J^P$. As in the case of the tetraquark $T_{cc}$, similar procedure applies to the tetraquark $T_{bb}$, which gives rise to numerical masses, mixing weights and magnetic moments of the doubly bottom systems. The computed results for the CMI matrices and the whole Hamiltonian are shown in Table \ref{tab:c2}. The resulted masses, mixing weights and magnetic moments of the $T_{bb}$ based on diagonalization are listed in Table \ref{tab:5} for the nosntrange states, in Table \ref{tab:6} for the states with strangeness $S=-1$, and \ref{tab:7} for that with strangeness $S=-2$.

\renewcommand{\tabcolsep}{0.55cm}
\renewcommand{\arraystretch}{1.2}
\begin{table*}[!htbp]
	\caption{Computed matrices of the CMI and the whole Hamiltonian of the doubly bottom tetraquarks $bb\bar{n}\bar{n}$, $bb\bar{s}\bar{n}$, and $bb\bar{s}\bar{s}$.}
	\label{tab:c2}
	\begin{tabular}{crcc}
		\bottomrule[1.0pt]\bottomrule[0.5pt]
		Systems             &$I(J^P)$     		& CMI matrices	& Hamiltonians	\\\hline
		$bb\bar{n}\bar{n}$  &$0(1^+)$
&$\begin{bmatrix}-140.7&-23.9\\-23.9&-16.6\\\end{bmatrix}$
&$\begin{bmatrix}-424.1&-23.9\\-23.9&125.1\\\end{bmatrix}$\\
                    		&$1(0^+)$
&$\begin{bmatrix}31.2&41.4\\41.4&80.6\\\end{bmatrix}$
&$\begin{bmatrix}-252.2&41.4\\41.4&222.4\\\end{bmatrix}$\\
                    		&$1(1^+)$       	&$42.5$	&$-240.9$				\\
                    		&$1(2^+)$        	&$65.0$	&$-218.4$				\\
        $bb\bar{s}\bar{n}$  &$1/2(0^+)$
&$\begin{bmatrix}14.9&42.9\\42.9&57.5\\\end{bmatrix}$
&$\begin{bmatrix}-359.6&42.9\\42.9&-28.6\\\end{bmatrix}$\\
                    		&$1/2(1^+)$
&$\begin{bmatrix}26.6&0.6&1.3\\0.6&-94.3&-24.8\\1.3&-24.8&-8.9\\\end{bmatrix}$
&$\begin{bmatrix}-347.9&0.6&1.3\\0.6&-468.8&-24.8\\1.3&-24.8&-94.9\\\end{bmatrix}$\\
                    		&$1/2(2^+)$        	&$50.0$	&$-324.5$				\\
		$bb\bar{s}\bar{s}$  &$0(0^+)$
&$\begin{bmatrix}8.6&44.5\\44.5&49.3\\\end{bmatrix}$
&$\begin{bmatrix}-457.0&44.5\\44.5&-264.4\\\end{bmatrix}$\\
                    		&$0(1^+)$       	&$20.7$	&$-444.8$				\\
                    		&$0(2^+)$        	&$45.0$	&$-420.6$				\\
		\bottomrule[0.5pt]\bottomrule[1.0pt]
	\end{tabular}
\end{table*}

\renewcommand{\tabcolsep}{0.55cm}
\renewcommand{\arraystretch}{1.2}
\begin{table*}[!htbp]
\caption{Masses (in MeV), mixing weights(for the rep. $\bar{3}_c\otimes3_c$) and magnetic moments (for $I=0$ or $I_{z}=1,0,-1$, in $\mu_N$) of the non-strange $T_{bb}$ states.  $dM$ is the mass spacing relative to the two-meson threshold.}
	\label{tab:5}
	\begin{tabular}{cclccc}
	\bottomrule[1.0pt]\bottomrule[0.5pt]
$I$  &$J^P$	&Masses	(MeV)	&$dM$	&Mixings ($\%$)& Magnetic moments ($I_{z}$)	\\\hline
		$0$     		       	&$1^+$	&$10406\pm17$	        &$-198$		&$100$		&$-0.13$			\\
								&		&$10957\pm18$			&$353$		&			&$-0.86$			\\
		$1$		            	&$0^+$	&$10576\pm17$			&$17$		&$99$		&					\\
								&		&$11057\pm18$			&$498$		&$1$		&					\\
								&$1^+$	&$10590\pm17$			&$-14$		&$100$		&$0.80,-0.49,-1.79$	\\
								&$2^+$	&$10613\pm17$	        &$-37$		&$100$		&$1.61,-0.99,-3.58$	\\		
		\bottomrule[0.5pt]\bottomrule[1.0pt]
	\end{tabular}
\end{table*}

\renewcommand{\tabcolsep}{0.53cm}
\renewcommand{\arraystretch}{1.2}
\begin{table*}[!htbp]
	\caption{Masses (in MeV), mixing weights of the rep. $\bar{3}_c\otimes3_c$ and magnetic moments (for $I_{z}=1/2,-1/2$, in $\mu_N$) of the tetraquark $T_{bb,s}$ with strangeness $S=-1$. $dM$ is the mass spacing relative to the two-meson threshold. $M(BB^*_s)=10695$ MeV.}
	\label{tab:6}
	\begin{tabular}{cclccc}
		\bottomrule[1.0pt]\bottomrule[0.5pt]
		$I$            	&$J^P$			&Masses	(MeV)		      &$dM$	        &Mixings ($\%$)&Magnetic moments ($I_{z}$)	\\\hline
		$1/2$  		 	&$0^+$  		&$10652\pm26$		      &$5$		    &$98$		&					\\
						&				&$10994\pm30$		      &$347$		&$2$		&					\\
						&$1^+$  		&$10547\pm7$        	  &$-148$	    &$100$		&$-0.12,-0.11$		\\
						&				&$10669\pm26$		      &$-26$		&$100$		&$0.66,-0.66$		\\
						&				&$10924\pm15$		      &$229$		&			&$1.43,-1.15$		\\
						&$2^+$  		&$10693\pm26$       	  &$-47$		&$100$		&$1.31,-1.28$		\\		
		\bottomrule[0.5pt]\bottomrule[1.0pt]
	\end{tabular}
\end{table*}

\textsl{Nonstrange tetrarquarks} In the case of non-strange $T_{bb}$, hadron mass lies in the range from $10.4$ to $11.1$ GeV, with gap of about $0.65$ GeV. In contrast with the doubly charmed tetraquarks, all of which are above their relevant two-meson thresholds, some of the doubly bottom tetraquark $T_{bb}$ binds more strongly to lie below their relevant two-meson thresholds. The typical examples are the $T_{bb}(10406,01^+)^-$ and $T_{bb}(10613,12^+)$, as well as the strange tetraquarks $T_{bb,s}(10547,(1/2)1^+)$, $T_{bb,s}(10693,(1/2)2^+)$, and $T_{bb,ss}(10783,02^+)^0$, which bound deeply and are in the color rep. $\bar{3}_c\otimes3_c$. This can be due to the enhanced deep binding $B_{bb}$ between the bottom quarks \cite{Karliner:2017qjm} compared to the binding $B_{cc}$ between two charm quarks. As shown in Table \ref{tab:5}, the binding is more deep for the isoscalar tetraquark $T_{bb}(10406,01^+)^-$, 198 MeV below the $BB^*$ threshold, and make it more stable against the strong decays. As shown in Table \ref{tab:BB}, we have compared the predicted mass of this state with the results from other works.

\renewcommand{\tabcolsep}{0.55cm}
\renewcommand{\arraystretch}{1.2}
\begin{table*}[!htbp]
	\caption{Masses (in MeV), mixing weights of the rep. $\bar{3}_c\otimes3_c$ and magnetic moments (for $I=0$, in $\mu_N$) of the tetraquark $T_{bb,ss}$ with strangeness $S=-2$. $dM$ is the mass spacing relative to the two-meson threshold.}
	\label{tab:7}
	\begin{tabular}{cclccc}
		\bottomrule[1.0pt]\bottomrule[0.5pt]
			$I$            			&$J^P$			&Masses (MeV)	&$dM$   	&Mixings ($\%$)&Magnetic moments ($I_{z}$)	\\\hline
			$0$			  	  		&$0^+$  		&$10736\pm25$	&$2$		&$95$		&					\\
									&				&$10949\pm28$	&$215$		&$5$		&					\\
									&$1^+$			&$10758\pm24$	&$-24$		&$100$		&$0.51$				\\
									&$2^+$			&$10783\pm24$ 	&$-48$		&$100$		&$1.02$				\\	
		\bottomrule[0.5pt]\bottomrule[1.0pt]
	\end{tabular}
\end{table*}

\renewcommand{\tabcolsep}{0.55cm}
\renewcommand{\arraystretch}{1.2}
\begin{table*}[!htbp]
	\caption{ Comparison of predictions for $M(T^-_{bb})$. $M(BB^*)=10604$ MeV. $M(BB\pi)=10699$ MeV.}
	\label{tab:BB}
	\begin{tabular}{lcc}
		\bottomrule[1.0pt]\bottomrule[0.5pt]
		References                          	& Masses (MeV)							& Method\\ \hline
		Present work                         	&$10406\pm17$                           &\\
		\cite{Zhang:2021yul}					&$10654$								& MIT bag model\\
		\cite{Liu:2023vrk}						&$10298$								& Chromomagnetic interactions model\\
		\cite{Luo:2017eub}						&$10686$								& same as above\\
		\cite{Karliner:2017qjm}                 &$10389\pm12$							& Mass sum relation\\
		\cite{Maiani:2022qze}	                &$10552$								& BO bound \\
		\cite{Maiani:2019lpu}					&$10466(10448)$							& Hydrogen bond of QCD\\
		\cite{Eichten:2017ffp}					&$10482$								& Heavy-quark symmetry relations\\
		\cite{Song:2023izj}						&$10530$								& Regge relation + mass scaling\\
		\cite{Ebert:2007rn}						&$10502$								& Relativistic quark model\\
		\cite{Braaten:2020nwp}					&$10471\pm25$							& Color-Coulomb potential + heavy-quark limit\\
		\cite{Lu:2020rog}						&$10550$								& Relativistic quark model\\
		\cite{Navarra:2007yw}					&$10200\pm300$							& QCD sum rule\\
		\cite{Vijande:2006jf}					&$10426$								& Constituent quark model\\
		\cite{Vijande:2003ki}					&$10261$								& Chiral constituent quark model\\
		\bottomrule[0.5pt]\bottomrule[1.0pt]
	\end{tabular}
\end{table*}

To estimate lifetime of the $T_{bb}$, we focus on the charged vertex $b\to cW^{*-}$, where the $W^{*-}$ couples to the products $e\nu$, $\mu\nu$, $\tau\nu$, or the color charges of $d\bar{u}$ and $s\bar{c}$ \cite{Karliner:2014gca}. We consider the initial state to the lowest state $T_{bb}(10406,01^{+})$, so that $M_i=10406$ MeV. The final state is $BD$, with the mass $M_f=M_B+M_D=7147$ MeV.  Applying Eq. (\ref{for:Gan}) to the $T_{bb}(10406,01^{+})$ decaying to $BD$ gives the weak decay width
\begin{equation}\label{for:Ga}
	\begin{aligned}
        \Gamma(T_{bb}(10406,01^+)^-)=\frac{18G_F^2M_i^5}{192\pi^3}F(x_{BD})|V_{cb}|^2=2.02\times10^{-12}\ \text{GeV},
	\end{aligned}
\end{equation}
where the matrix element $|V_{cb}|=0.042$ \cite{ParticleDataGroup:2024cfk}, the Fermi constant $G_F=1.166\times 10^{-5}$ GeV$^{-2}$ \cite{ParticleDataGroup:2024cfk}, $x_{BD}=(M_f/M_i)^2=0.472$ and $F(x_{BD})=0.023$ by Eq. (\ref{for:ksf}). This amounts to  lifetime of $326$ fs and is consistent with the predicted results from Refs. \cite{Karliner:2017qjm,Gao:2020bvl}.

To find the dominant decay channels, we consider the following ways of the weak decays:
\begin{itemize}
	\item [i)] Weak processes involving the charged vertex $b\to cW^{*-}$. In this situation, the processes $T_{bb}(10406,01^+)^-\to \bar{B}^0D^0\pi^-$ or $B^-D^+\pi^-$ are more likely to occur, suggesting searching for the $T^-_{bb}$ structure in these two channels. Other relevant processes contain the decay processes $T_{bb}(10406,01^+)^-\to J/\Psi B^-\bar{K}^0$, $J/\Psi \bar{B}^0K^-$, $J/\Psi J/\Psi K^-\bar{K}^0$ and $D^0 D^+\pi^-\pi^-$, and the estimates of their branching fractions will not exceed that of the aforementioned two processes.

	\item [ii)] Alternative channel involving the scattering between $\bar{d}$ and $b$ quarks, which is associated with the mesons $B/D$. In the heavy-light systems, however, the spatial wave function $lim_{x_i\to x_j}|\Psi_{ij}(x_i-x_j)|^2$ for overlap of these two quarks is quite small, thus the decay channels $T_{bb}(10406,01^+)^-\to J/\Psi B^-$ and $B^-D^0$ are hard to happen and can not be the dominant decay channels.
\end{itemize}

\textsl{Strange tetraquark} In the case of the strange state $T_{bb,s}$, the mass range is from $10.5$ to $11.0$ GeV, with a mass gap of $0.45$ GeV; whereas for the strange $T_{bb,ss}$, the mass range is from $10.7$ to $10.9$ GeV, with a mass gap of $0.21$ GeV. Evidently, as the number of strange quarks increases, the mass gap of the $T_{bb}$ system becomes significantly narrower. Among these strange states, there is an analogue of the state $T_{bb}(10406)^-$, denoted as $T_{bb,s}(10547,(1/2)1^+)$, which lies below the $BB^*_s$ threshold about 148 MeV and is stable against both of the strong and electromagnetic decays. It can decay via the standard weak process $b\to cW^{*-}$ to the three color singlets of $\bar{c}s$ and $\bar{u}d$, and $e\nu$, $\mu\nu$, and $\tau\nu$.

Assuming the initial state mass to be $M_i=10547$ MeV, the final state mass to be $M_f=M(B_s)+M(D)=7234$ MeV, which gives the kinematic suppression factor $x_{B_sD}=0.47$, one can estimate the total width to be
\begin{equation}\label{for:Ga5}
	\begin{aligned}
        \Gamma(T_{bb,s}(10547,(1/2)1^+))=\frac{18G_F^2M_i^5}{192\pi^3}F(x_{B_sD})|V_{cb}|^2=2.19\times10^{-12}\ \text{GeV},
	\end{aligned}
\end{equation}
with the factor of 2 accounting for the two bottom quarks. From this, we can roughly estimate its lifetime, which is 301 fs.
Here, we have taken the small CKM matrix element $V_{ub}=0$, which does not significantly shorten the predicted lifetime of $T^-_{bb,s}$.

\subsection{The bottom-charmed states}
\label{sec:bc}

\renewcommand{\tabcolsep}{0.03cm}
\renewcommand{\arraystretch}{1.2}
\begin{table*}[!htbp]
	\caption{The computed matrices of the CMI and the whole Hamiltonian of the bottom-charmed tetraquarks $bc\bar{n}\bar{n}$, $bc\bar{s}\bar{n}$, and $bc\bar{s}\bar{s}$.}
	\label{tab:c3}
	\begin{tabular}{crcc}
		\bottomrule[1.0pt]\bottomrule[0.5pt]
		Systems             &$I(J^P)$     		& CMI matrices	& Hamiltonians	\\\hline
		$bc\bar{n}\bar{n}$  &$0(0^+)$
&$\begin{bmatrix}-160.0 & 85.9 \\85.9 & -143.6\end{bmatrix}$
&$\begin{bmatrix}-336.4 & 85.9 \\85.9 & -55.4\end{bmatrix}$\\
		                    &$0(1^+)$
&$\begin{bmatrix}-141.2 & 36.4 & -49.6 \\36.4 & -85.1 & 42.9 \\-49.6 & 42.9 & -17.2\end{bmatrix}$
&$\begin{bmatrix}-317.5 & 36.4 & -49.6 \\36.4 & 3.1 & 42.9 \\-49.6 & 42.9 & 71.0\end{bmatrix}$\\
                            &$0(2^+)$           &$31.8$                    &$120.0$\\
                    		&$1(0^+)$
&$\begin{bmatrix}6.6 & 85.9 \\85.9 & 80.0\end{bmatrix}$
&$\begin{bmatrix}-169.8 & 85.9 \\85.9 & 168.2\end{bmatrix}$\\
                    		&$1(1^+)$       	
&$\begin{bmatrix}30.0 & 17.1 & 36.4 \\17.1 & 34.5 & -49.6 \\36.4 & -49.6 & 70.6\end{bmatrix}$
&$\begin{bmatrix}-146.4 & 17.1 & 36.4 \\17.1 & -141.9 & -49.6 \\36.4 & -49.6 & 158.8\end{bmatrix}$\\
                    		&$1(2^+)$        	&$76.7$	&$-99.7$				\\
        $bc\bar{s}\bar{n}$  &$1/2(0^+)$
&$\begin{bmatrix}-9.5 & -0.2 & 0.4 & 87.1 \\-0.2 & -113.6 & 87.1 & 0.0 \\0.4 & 87.1 & -137.5 & -0.4 \\87.1 & 0.0 & -0.4 & 56.8\end{bmatrix}$
&$\begin{bmatrix}-269.5 & -0.2 & 16.4 & 87.1 \\-0.2 & -373.5 & 87.1 & 16.0 \\16.4 & 87.1 & -258.1 & -0.4 \\87.1 & 16.0 & -0.4 & -63.8\end{bmatrix}$\\
                    		&$1/2(1^+)$
&$\begin{bmatrix}14.2 & 0.5 & 17.0 & 0.2 & 36.1 & 1.0 \\0.5 & -94.8 & 0.1 & 36.1 & 0.0 & -50.3 \\17.0 & 0.1 & 19.0 & 0.9 & -50.3 & 0.0 \\0.2 & 36.1 & 0.9 & -78.2 & 1.2 & 42.5 \\36.1 & 0.0 & -50.3 & 1.2 & 47.4 & 0.2 \\1.0 & -50.3 & 0.0 & 42.5 & 0.2 & -9.5\end{bmatrix}$
&$\begin{bmatrix}-245.8 & 0.5 & 17.0 & 16.2 & 36.1 & 1.0 \\0.5 & -354.7 & 0.1 & 36.1 & 16.0 & -50.3 \\17.0 & 0.1 & -240.9 & 0.9 & -50.3 & 16.0 \\16.2 & 36.1 & 0.9 & -198.9 & 1.2 & 42.5 \\36.1 & 16.0 & -50.3 & 1.2 & -73.3 & 0.2 \\1.0 & -50.3 & 16.0 & 42.5 & 0.2 & -130.2\end{bmatrix}$\\
                    		&$1/2(2^+)$
&$\begin{bmatrix}61.6 & -0.2 \\-0.2 & 40.3\end{bmatrix}$
&$\begin{bmatrix}-198.3 & 15.8 \\15.8 & -80.3\end{bmatrix}$\\
		$bc\bar{s}\bar{s}$  &$0(0^+)$
&$\begin{bmatrix}-15.7 & 88.3 \\88.3 & 48.6\end{bmatrix}$
&$\begin{bmatrix}-359.1 & 88.3 \\88.3 & -280.8\end{bmatrix}$\\
                    		&$0(1^+)$
&$\begin{bmatrix}8.4 & 16.9 & 35.8 \\16.9 & 13.6 & -51.0 \\35.8 & -51.0 & 39.2\end{bmatrix}$
&$\begin{bmatrix}-335.1 & 16.9 & 35.8 \\16.9 & -329.9 & -51.0 \\35.8 & -51.0 & -290.3\end{bmatrix}$\\
                    		&$0(2^+)$        	&$56.5$	&$-287.0$				\\
		\bottomrule[0.5pt]\bottomrule[1.0pt]
	\end{tabular}
\end{table*}

\renewcommand{\tabcolsep}{0.55cm}
\renewcommand{\arraystretch}{1.2}
\begin{table*}[!htbp]
	\caption{Masses (in MeV), mixing weights(for the rep. $\bar{3}_c\otimes3_c$) and magnetic moments (for $I=0$ or $I_{z}=1,0,-1$, in $\mu_N$) of the non-strange $T_{cb}$ states.  $dM$ is the mass spacing relative to the two-meson threshold. $M(B^*D)=7192$ MeV.}
	\label{tab:4.1}
	\begin{tabular}{cclccc}
		\bottomrule[1.0pt]\bottomrule[0.5pt]
		$I$            			&$J^P$				&Masses	(MeV)	&$dM$	    &Mixings ($\%$)	&Magnetic moments ($I_{z}$)		\\\hline
		$0$		  		  		&$0^+$  			&$7142\pm13$	&$-5$		&$93$			&						\\
								&	  				&$7471\pm14$	&$324$		&$7$			&						\\
								&$1^+$  			&$7173\pm16$    &$-19$      &$97$			&$0.26$					\\
								&	  				&$7494\pm15$	&$302$		&$3$			&$-0.56$				\\
								&	  				&$7596\pm16$	&$404$		&$1$			&$-0.54$				\\
								&$2^+$  			&$7622\pm14$	&$289$		&				&$-0.56$				\\
		$1$		  		  		&$0^+$  			&$7312\pm16$	&$165$		&$95$			&						\\
								&		  			&$7691\pm17$	&$544$		&$5$			&						\\
								&$1^+$  			&$7330\pm14$	&$138$		&$97$			&$1.04,-0.83,-2.69$		\\
								&		  			&$7375\pm14$	&$183$		&$100$			&$1.67,-0.28,-2.23$		\\
								&		  			&$7672\pm14$	&$480$		&$3$			&$0.33,0.26,0.19$		\\
								&$2^+$  			&$7402\pm16$	&$69$		&$100$			&$2.03,-0.56,-3.16$		\\
		\bottomrule[0.5pt]\bottomrule[1.0pt]
	\end{tabular}
\end{table*}

Now, let us consider the bottom-charmed tetraquarks, denoted as $T_{cb}(M,IJ^P)$ when it has the mass $M$ and quantum number $IJ^P$. The procedure stated in Sect. \ref{sec:for} applies for exploring the bottom-charmed tetraquarks $T_{cb}$. As such, having the CMI matrices and the whole Hamiltonian for them computed, as shown in Table \ref{tab:c3}, one can compute numerical masses, mixing weights and magnetic moments of these doubly bottom systems, with the results listed in Tables \ref{tab:4.1}, \ref{tab:4.2}, and \ref{tab:4.21}. As we see, the bottom-charmed tetraquark mass ranges in the region of $7.14$\text{-}$7.69$ GeV for the nonstrange $T_{cb}$ states, in the region of $7.37$\text{-}$7.86$ GeV for the strange $T_{cb,s}$ states, and in the region of $7.66$\text{-}$8.03$ GeV for the doubly-strange $T_{cb,ss}$ states.

\renewcommand{\tabcolsep}{0.52cm}
\renewcommand{\arraystretch}{1.2}
\begin{table*}[!htbp]
	\caption{Masses (in MeV), mixing weights of the rep. $\bar{3}_c\otimes3_c$ and magnetic moments (for $I_{z}=1/2,-1/2$, in $\mu_N$) of the tetraquark $T_{cb,s}$ with strangeness $S=-1$. $dM$ is the mass spacing relative to the two-meson threshold. $M(BD_s)=7248$ MeV. $M(BD^*_s)=7392$ MeV. $M(B^*D^*_s)=7437$ MeV.}
	\label{tab:4.2}
	\begin{tabular}{cclccc}
		\bottomrule[1.0pt]\bottomrule[0.5pt]
		$I$            	&$J^P$	&Masses	(MeV)		&$dM$	    &Mixing ($\%$)&Magnetic moments	($I_{z}$)\\\hline
		$1/2$			&$0^+$	&$7266\pm8$    		&$18$		&$77$		&					\\
						&		&$7387\pm24$		&$139$		&$89$		&					\\
						&		&$7477\pm8$	       	&$229$		&$22$		&					\\
						&		&$7657\pm28$		&$409$		&$12$		&					\\
						&$1^+$	&$7310\pm7$		    &$-82$		&$88$		&$0.34,0.33$		\\
						&		&$7409\pm22$		&$17$		&$92$		&$0.83,-1.17$		\\
						&		&$7457\pm20$		&$65$		&$92$		&$1.44,-0.66$		\\
						&		&$7490\pm11$		&$98$		&$16$		&$0.72,-0.58$		\\
						&		&$7584\pm13$		&$192$		&$3$		&$1.53,-0.73$		\\
						&		&$7635\pm26$		&$243$		&$9$		&$0.36,0.24$		\\
						&$2^+$	&$7488\pm24$		&$51$		&$98$		&$1.74,-0.86$		\\
						&		&$7610\pm12$		&$173$		&$2$		&$1.74,-0.86$		\\		
		\bottomrule[0.5pt]\bottomrule[1.0pt]
	\end{tabular}
\end{table*}

\begin{figure}[htpb]
\begin{center}
\vspace*{0em}
\hspace*{0em}
\includegraphics[scale=0.38]{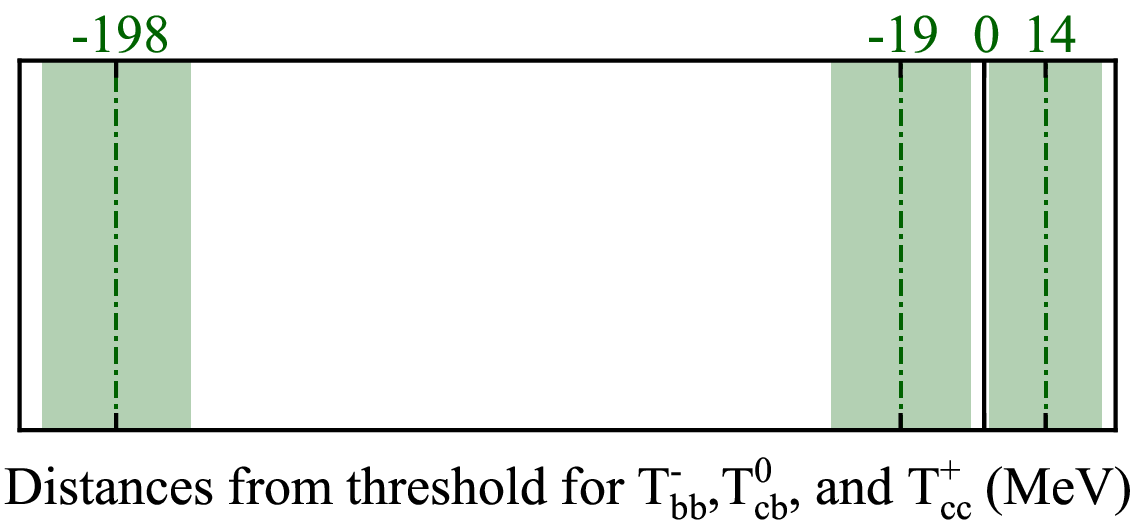}
\caption{Distances relative to the two-meson thresholds with $BB^*$, $DB^*$, and $DD^*$ are shown for the $T^-_{bb}$, $T^0_{cb}$, and $T^+_{cc}$ states (green lines).}\label{T}
\end{center}
\end{figure}

\textsl{Nonstrange tetraquarks}  As shown in Tables \ref{tab:4.1}, most of the tetraquark $T_{cb}$ unbind as they lie above the two-meson threshold $M(B^{*}D)=7192$ MeV. Two exceptions are the $00^{+}$ state at $7142$ MeV and the $01^{+}$ state at $7173$ MeV, which are likely to be a loosely bound state similar to the $T_{cc}$ states. The mass spacing $dM=M(T_{cb})-M(B^{(*)}D)$ of two states relative to $B^{(*)}D$ are
\begin{equation}\label{for:519}
	\begin{aligned}
		dM(T_{cb},00^{+})=-5\pm13\ \text{MeV},\ dM(T_{cb},01^{+})=-19\pm16\ \text{MeV},
	\end{aligned}
\end{equation}
where the uncertainty stems largely from that due to the model parameters.

Our computations imply that the binding depths of the $T_{cc}$, $T_{cb}$, and $T_{bb}$ obey $T_{cc}<T_{cb}<T_{bb}$ in size(see FIG. \ref{T}), being consistent with the values of the binding energy $B_{QQ^\prime}$ in Eq. (\ref{for:Bdi}) and Eq. (\ref{for:BQQ}). One anticipates then the state $T_{cb}(7173,01^+)^0$ with a mass of $7173\pm16$ MeV to be a state below the $B^*D$ mass threshold. For the below-threshold binding of the $T_{cb}$ and the $T_{bb}$, one may expect that two heavy quarks form a core with size of about $0.3$ fm or below, quite smaller than the size of hadrons and different in size depending on the flavor contents \cite{Eichten:2017ffp}. It is also suggested \cite{Karliner:2017qjm} that the interquark potential is singular at the origin, leading to a rapid increase in the $QQ^\prime$ binding energy as the distance decreases.

\renewcommand{\tabcolsep}{0.55cm}
\renewcommand{\arraystretch}{1.2}
\begin{table*}[!htbp]
	\caption{Masses (in MeV), mixing weights of the rep. $\bar{3}_c\otimes3_c$ and magnetic moments(for $I=0$, in $\mu_N$) of the tetraquark $T_{cb,ss}$ with strangeness $S=-2$. $dM$ is the mass spacing relative to the two-meson threshold. $M(B_sD^*_s)=7479$ MeV.}
	\label{tab:4.21}
	\begin{tabular}{cclccc}
		\bottomrule[1.0pt]\bottomrule[0.5pt]
		$I$            			&$J^P$				&Masses	(MeV)	&$dM$	&Mixings ($\%$)&Magnetic moments ($I_{z}$)		\\\hline
		$0$		  		  		&$0^+$  			&$7457\pm24$	&$122$		&$70$		&						\\
								&	  				&$7651\pm26$	&$316$		&$30$		&						\\
								&$1^+$  			&$7486\pm21$	&$7$		&$71$		&$0.56$					\\
								&	  				&$7557\pm21$	&$78$		&$99$		&$1.16$					\\
								&	  				&$7624\pm23$	&$145$		&$30$		&$0.45$					\\
								&$2^+$  			&$7587\pm23$	&$59$		&$100$		&$1.44$					\\
		\bottomrule[0.5pt]\bottomrule[1.0pt]
	\end{tabular}
\end{table*}

To estimate the lifetime of the loosely bound state $T_{cb}(7173,01^+)^0$ in Table \ref{tab:4.1}, one has to consider, beside the contributions from weak decays, the contributions from electromagnetic transition. The initial state $T_{cb}(7173,1^+)^0$ decays into the final state $T_{cb}(0^+)^0\gamma$, assuming the $T_{cb}^0$ to have a zero lifetime (see Eq. (\ref{for:519}), and the $0(0^+)$ state is more loosely bound than the $0(1^+)$ state).
For the state $T_{cb}(7173,1^+)^0$, one can first extract the mixing coefficients directly from the solved eigenvectors with respect to the eigenvalue 7173 (MeV) in Table \ref{tab:c3} for the Hamiltonian of the system $bc\bar{n}\bar{n}$ with the quantum number $0(1^+)$. The numerical results for these coefficients (in Table \ref{tab:4.1}, we omit them and all others for concise) are
\begin{equation}\label{for:R26}
	\begin{aligned}
   (R_{26},R_{13})&=(-0.96, 0.27),\\
   (R_{24},R_{12},R_{15})&=(0.98, -0.13, 0.14).
	\end{aligned}
\end{equation}

As stated  at the beginning of Sect. \ref{sec:CD}, one can employ the wavefunction of the $T_{cb}(7173,1^+)^0$ to derive the transition magnetic moment
\begin{equation}\label{for:tr}
	\begin{aligned}
		\mu &\left[T_{cb}(7173,1^+)^0\to T_{cb}(0^+)^0\right]=\left\langle\psi^{J=0,J_z=0}_2|\hat{\mu}_{\text{TOT}}|\psi^{J=1,J_z=0}_5\right\rangle\\
                    &=R_{26}R_{24}(\mu_c-\mu_b)+\frac{R_{13}R_{15}}{\sqrt{3}}(\mu_b-\mu_c)+\frac{R_{13}R_{12}}{\sqrt{3}}\sqrt{2}(\mu_b+\mu_c-\mu_{\bar{u}}-\mu_{\bar{d}})\\
                    &=-0.443.
	\end{aligned}
\end{equation}
Given the initial and final state masses $M_i=7173$ MeV and $M_f=M(B)+M(D)=7147$ MeV, the momentum of the outgoing photon $\omega=26$ MeV, as well as other parameters given above, one can apply Eq. (\ref{for:Ga2}) to get the M1 decay width
\begin{equation}\label{for:Ga71}
	\begin{aligned}
		\Gamma\left(T_{cb}(7173,1^+)^0\to T_{cb}(0^+)^0\gamma\right)=9.58\times10^{-9}\ \text{GeV}.
	\end{aligned}
\end{equation}

Next, we consider the weak decays associated with the charged vertex $c\to sW^{*+}$, where $W^{*+}$ couples to $e\nu$, $\mu\nu$, and three colors of $u\bar{d}$. In this situation, the transitions $T_{cb}(7173,01^+)^0\to B^-\bar{K}^0\pi^+$ and $\bar{B}^0K^-\pi^+$ are more likely to occur, which suggests to search for the $T^0_{cb}$ in these two channels. Using the initial and final state masses $M_i=7173$ MeV and $M_f=M(B)+M(K)=5775$ MeV, the squared ratio $x_{BK}=0.648$ of two masses, and the CKM matrix element $|V_{cs}|=0.973$ \cite{ParticleDataGroup:2024cfk}, one has
\begin{equation}\label{for:Gax}
	\begin{aligned}
		\Gamma(T_{cb}(7173,01^+)^0)=\frac{5G_F^2M_i^5}{192\pi^3}F(x_{BK})|V_{cs}|^2=5.40\times10^{-12}\ \text{GeV}.
	\end{aligned}
\end{equation}
The two contributions from Eqs. (\ref{for:Ga71}) and (\ref{for:Gax}) amount to the lifetime of $T_{cb}(7173,01^+)^0$, which is 0.069 fs. This lifetime is much shorter than that of the states $T_{bb}(10406,01^+)^-$ and $T_{bb,s}(10547,(1/2)1^+)$. Note that for states that allow electromagnetic decays, the contribution of weak decay processes to lifetime becomes negligible (with four orders of magnitude smaller). Moreover, the decays associated with charged vertices which we do not consider yet include that associated with transitions $b\to cW^{*-}$, $c\to dW^{*+}$ and $bc\to cs$. The later transitions correspond to the channels like $D^0\bar{K}^0$, $D^+K^-$ and $\bar{B}^0\pi^+\pi^-$, which contribute negligibly to the lifetime of the $T_{cb}$ state.

There are some multiplets, such as $T_{cb}(7330,11^+)$ and $T_{cb}(7375,11^+)$, which are near each other in weight. The ratios of the strong-decay partial widths of them with respect to different channels are
\begin{equation}\label{7330}
	\begin{aligned}
\frac{\Gamma(T_{cb}(7330,11^+)\to BD^*)}{\Gamma(T_{cb}(7330,11^+)\to B^*D)}=\frac{0.09\gamma_{BD^*}\alpha}{194.75\gamma_{B^*D}\alpha}=0.0005,
	\end{aligned}
\end{equation}
\begin{equation}\label{7375}
	\begin{aligned}
		\frac{\Gamma(T_{cb}(7375,11^+)\to BD^*)}{\Gamma(T_{cb}(7375,11^+)\to B^*D)}=\frac{129.32\gamma_{BD^*}\alpha}{3.33\gamma_{B^*D}\alpha}=38.83.
	\end{aligned}
\end{equation}
Evidently, they differ significantly, with three order of magnitude, and are then easy to be distinguished in experiments. We suggest to search these two states in the decay channels of the $BD^*$ and $B^*D$.

\textsl{Strange tetraquarks} For the case of the strange tetraquark $T_{cb,s}$ and $T_{cb,ss}$, the masses and magnetic moments are computable similarly and they are numerically listed in Table \ref{tab:4.2} and \ref{tab:4.21}, respectively. In the case of the $T_{cb,s}$, only one state, $T_{cb,s}(7310,(1/2)1^+)$, remains to be stable against strong decays, lying 82 MeV below the $BD^{*}_s$ mass threshold. In the case of the $T_{cb,ss}$, all states are above their respective mass thresholds.

To estimate the lifetime of the $T_{cb,s}(7310,(1/2)1^+)$ state, we consider the radiative transition $T_{cb,s}(7310,1^+)^+\to T_{cb,s}(0^+)^+\gamma$, and compute the corresponding decay widths as done for the $T_{cb}(7173)^0$. For the transition magnetic moment used in radiative decays, one can rewrite the third formula in Eq. (\ref{for:Ga2}) to get
\begin{equation}\label{for:125}
	\begin{aligned}
        \mu&=\left\langle\psi^{J=0,J_z=0}_3|\hat{\mu}_{\text{TOT}}|\psi^{J=1,J_z=0}_6\right\rangle\\
        &=\frac{R_{13}}{\sqrt{3}}\left(R_{14}\mu_{\bar{n}}-R_{14}\mu_{\bar{s}}+R_{15}\mu_b-R_{15}\mu_c\right)+\frac{R_{23}}{\sqrt{3}}\left(R_{24}\mu_{\bar{n}}-R_{24}\mu_{\bar{s}} +R_{25}\mu_b-R_{25}\mu_c\right)\\
        &+R_{16}\left(R_{14}\mu_c+R_{15}\mu_{\bar{s}}-R_{14}\mu_b-R_{15}\mu_{\bar{n}}\right)+R_{26}\left(R_{24}\mu_c+R_{25}\mu_{\bar{s}}-R_{24}\mu_b-R_{25}\mu_{\bar{n}}\right)\\
        &+\frac{R_{12}R_{13}+R_{22}R_{23}}{\sqrt{3}}\sqrt{2}(\mu_b+\mu_c-\mu_{\bar{n}}-\mu_{\bar{s}}).\\
	\end{aligned}
\end{equation}
For the $T_{cb,s}(7310,1^+)^+$ in Table \ref{tab:4.2}, the mixing coefficients are available from the solved eigenvectors with respect to the eigenvalue 7310 (MeV) in Table \ref{tab:c3} for the Hamiltonian of the system $bc\bar{s}\bar{n}$ with the quantum number $1/2(1^+)$. These coefficients (we omit in Table \ref{tab:4.2}) are
\begin{equation}\label{for:RIJ}
	\begin{aligned}
    (R_{23},R_{26},R_{13},R_{16})&=(-0.09,-0.87,0.47,0.06),\\
    (R_{22},R_{24},R_{25},R_{12},R_{14},R_{15})&=(0.05,0.93,-0.06,-0.25,-0.06,0.24).\\
	\end{aligned}
\end{equation}
Putting into Eq. (\ref{for:125}), one has for the transition magnetic moments for two isospin partners of the $T_{cb,s}(7310)$ state
\begin{equation}\label{for:135}
	\begin{aligned}
        \mu\left[T_{cb,s}(7310,1^+)^+\to T_{cb,s}(0^+)^+\right]&=-0.302,\\
        \mu\left[T_{cb,s}(7310,1^+)^0\to T_{cb,s}(0^+)^0\right]&=-0.229.
	\end{aligned}
\end{equation}
Given $M_i=7310$ MeV, $M_f=M(B)+M(D_s)=7248$ MeV, and $\omega=62$ MeV, we can apply Eq. (\ref{for:135}) to Eq. (\ref{for:Ga2}) to obtain the radiative widths \begin{equation}\label{}
	\begin{aligned}
		\Gamma\left(T_{cb,s}(7310,1^+)^+\to T_{cb,s}(0^+)^+\gamma\right)&=5.92\times10^{-8}\ \text{GeV},\\
        \Gamma\left(T_{cb,s}(7310,1^+)^0\to T_{cb,s}(0^+)^0\gamma\right)&=3.41\times10^{-8}\ \text{GeV},
	\end{aligned}
\end{equation}
which correspond to the lifetimes 0.011 fs for $T_{cb,s}(7310)^+$ and 0.019 fs for $T_{cb,s}(7310)^0$, with the contributions from weak decay ignored.

In the case of the strange $T_{cb,s}$, there are three multiplets having near masses 7409, 7457, and 7490 MeV. Similar computations give the partial width ratios for different channels shown below, which are,
\begin{equation}\label{for:7409}
	\begin{aligned}
		\frac{\Gamma(T_{cb,s}(7409,(1/2)1^+)\to BD^*_s)}{\Gamma(T_{cb,s}(7409,(1/2)1^+)\to B^*D_s)}=\frac{2.20\gamma_{BD^*_s}\alpha}{171.20\gamma_{B^*D_s}\alpha}=0.013,
	\end{aligned}
\end{equation}
\begin{equation}\label{for:7457}
	\begin{aligned}
		\frac{\Gamma(T_{cb,s}(7457,(1/2)1^+)\to BD^*_s)}{\Gamma(T_{cb,s}(7457,(1/2)1^+)\to B^*D_s)}=\frac{46.16\gamma_{BD^*_s}\alpha}{4.50\gamma_{B^*D_s}\alpha}=10.26,
	\end{aligned}
\end{equation}
\begin{equation}\label{for:7490}
	\begin{aligned}
		\frac{\Gamma(T_{cb,s}(7490,(1/2)1^+)\to BD^*_s)}{\Gamma(T_{cb,s}(7490,(1/2)1^+)\to B^*D_s)}=\frac{157.02\gamma_{BD^*_s}\alpha}{324.48\gamma_{B^*D_s}\alpha}=0.484.
	\end{aligned}
\end{equation}
For the $(B_sD)^*$ channel of the $T_{cb,s}(7409,(1/2)1^+)$, the $T_{cb,s}(7457,(1/2)1^+)$ and the $T_{cb,s}(7490,(1/2)1^+)$, the three ratios shown in Eqs. (\ref{for:7409})\text{-}(\ref{for:7490}) become 0.002, 1.907, and 0.472, respectively. These ratios differ evidently and are helpful for identifying or distinguishing these states in experiments.

\section{conclusions and remarks}
\label{sec:cr}
This work is devoted to study systematically the main static properties and inner-structure of nonstrange and strange tetraquarks with two heavy quarks within the picture of the QCD string with chromomagnetic interaction. The masses, magnetic moments and color-spin structures are computed for all DH (doubly charmed, doubly bottom and bottom-charmed) tetraquarks, and the hadron lifetime is estimated for those tetraquarks of doubly-bottom and bottom-charmed which lie below the relevant two-meson thresholds via analyzing the weak decay and the radiative decay. Two-body formula is applied to explore the relative strong decay widths of the doubly-charmed and bottom-charmed tetraquarks.

Our mass analysis indicates that all nonstrange doubly charmed tetraquarks $T_{cc}$ are above the relevant two-meson thresholds and are unstable against strong decays to two-meson dissociation. There are a relatively stable state among the doubly charmed tetraquarks with strangeness of -1 and -2, which is the lowest state, $T_{cc,s}(4034,(1/2)1^+)$, $55\pm31$ MeV above the $(D_sD)^*$ mass threshold. We anticipate that the strange tetraquark $T_{cc,s}(4034,(1/2)1^+)$ is more likely to be discovered in the channels of $T_{cc,s}\to D_sD^*$ and $D_s^*D$.
In contrast, some of the doubly-bottom tetraquark $T_{bb}$ binds more strongly to lie below their relevant two-meson thresholds, which are the nonstrange state $T_{bb}(10406,01^+)^-$ and the strange state $T_{bb,s}(10547,(1/2)1^+)$, all in the color rep. $\bar{3}_c\otimes3_c$. We find that two doubly-bottom states, $T_{bb}(10406,01^+)^-$ and $T_{bb,s}(10547,(1/2)1^+)$, are stable against the radiative decay and strong decay, having lifetimes of $326$ fs and $301$ fs, respectively.

In the bottom-charmed sector of the DH tetraquarks $T_{cb}$, there are two tetraquarks, a nonstrange state $T_{cb}(7173)^0$ and a strange state $T_{cb,s}(7310)$ found to lie below the $B^*D$ or $BD^*_s$ thresholds, stable thereby against strong interactions of two-meson dissociation. We analyze the decay behaviors of two tetraquarks via weak and electromagnetic interactions, and find that the nonstrange state $T_{cb}(7173)^0$ has a lifetime of 0.069 fs and the strange state $T_{cb,s}(7310)$ has the lifetimes of 0.011 fs and 0.019 fs, corresponding to its isospin multiples.

Two color-spin multiplets of the doubly charmed tetraquark $T_{cc}$ are predicted, which are the lower state $T_{cc}(3890.3, 01^{+})$ at 3890.3 MeV, with mixing weight $(R_1)^{2}=0.95$ for the exotic component $\bar{3}_c \otimes 3_c$, and the higher state $T^\prime_{cc}(4244.3,01^{+})$ at 4244.3 MeV. The lower one is in consistent with the LHCb-reported tetraquark $T_{cc}(3875)^+$ with $01^+$, near to the $DD^*$ mass threshold $M(DD^*)=3875.8$ MeV. Furthermore, two heavy-flavor counterparts of the $T_{cc}(3875)^+$, the $T^-_{bb}(01^+)$ and $T^0_{cb}(01^+)$, are found to lie below the two-meson thresholds. We suggest experimental searches for these two states in the decay channels $T_{bb}^-\to \bar{B}^0D^0\pi^-$ or $B^-D^+\pi^-$ and $T_{cb}^0\to B^-D^+\gamma$ or $\bar{B}^0D^0\gamma$, respectively.

The calculated chromomagneic mixing of the color-spin states of the DH tetraquarks implies that the mixing is deep for the doubly-charmed and bottom-charmed tetraquarks, especially for the strange case of these systems. We find that the magnetic moments are sensitive to chromomagneic mixing of the color-spin states of these tetraquarks, suggesting a valuable probe to experimentally detect the nontrivial color configuration $6_{c}\otimes\bar{6}_{c}$.

\medskip
\textbf{ACKNOWLEDGMENTS}

D. J thanks A. Hosaka for discussions. D. J. is supported by
the National Natural Science Foundation of China under the no. 12165017.

\appendix
\section{Wave functions of tetraquarks}
\label{sec:scw}

For numerical calculations of masses, mixing weights, and magnetic moments, it is essential to employ color-spin wavefunctions to describe the chromomagnetic structure of a given tetraquark state.
In the case of a ground-state tetraquark, its total angular momentum should be determined entirely by the intrinsic angular momentum of its constituent quarks.
For the spin configuration of the DH tetraquark $QQ\bar{q}\bar{q}$, the possible wavefunctions are
\begin{equation}\label{for:stetr}
	\begin{aligned}
		\chi^{2,2}_{1}&=\uparrow\uparrow\uparrow\uparrow,\\
		\chi^{1,1}_{2}&=\frac{1}{2}(\uparrow\uparrow\uparrow\downarrow+\uparrow\uparrow\downarrow\uparrow-\uparrow\downarrow\uparrow\uparrow-\downarrow\uparrow\uparrow\uparrow),\\
        \chi^{0,0}_{3}&=\frac{1}{\sqrt{3}}(\uparrow\uparrow\downarrow\downarrow+\downarrow\downarrow\uparrow\uparrow)-\frac{1}{2\sqrt{3}}(\uparrow\downarrow\uparrow\downarrow+\uparrow\downarrow\downarrow\uparrow+\downarrow\uparrow\uparrow\downarrow+\downarrow\uparrow\downarrow\uparrow),\\
		\chi^{1,1}_{4}&=\frac{1}{\sqrt{2}}(\uparrow\uparrow\uparrow\downarrow-\uparrow\uparrow\downarrow\uparrow),\\
		\chi^{1,1}_{5}&=\frac{1}{\sqrt{2}}(\uparrow\downarrow\uparrow\uparrow-\downarrow\uparrow\uparrow\uparrow),\\
        \chi^{0,0}_{6}&=\frac{1}{2}(\uparrow\downarrow\uparrow\downarrow-\uparrow\downarrow\downarrow\uparrow-\downarrow\uparrow\uparrow\downarrow+\downarrow\uparrow\downarrow\uparrow),\\
	\end{aligned}
\end{equation}
where $\uparrow$ and $\downarrow$ denote the $z$-components of the quark spins.

\renewcommand{\tabcolsep}{0.55cm}
\renewcommand{\arraystretch}{1.2}
\begin{table*}[!htbp]
	\caption{The wavefunction we constructed for the DH tetraquarks with given flavor, color, and spin configurations.}
	\label{tab:ml}
	\begin{tabular}{ll}
		\bottomrule[1.0pt]\bottomrule[0.5pt]
		$0^+$&$\psi^{0,0}_1=R_1\phi_2\chi^{0,0}_3+R_2\phi_1\chi^{0,0}_6$\\
		&$\psi^{0,0}_2=R_1\phi_2\chi^{0,0}_6+R_2\phi_1\chi^{0,0}_3$\\
		&$\psi^{0,0}_3=R_1\phi_2\chi^{0,0}_3+R_2\phi_2\chi^{0,0}_6+R_3\phi_1\chi^{0,0}_3+R_4\phi_1\chi^{0,0}_6$\\
		$1^+$&$\psi^{1,1}_1=R_1\phi_2\chi^{1,1}_2$\\
		&$\psi^{1,1}_2=R_1\phi_2\chi^{1,1}_4+R_2\phi_1\chi^{1,1}_5$\\
		&$\psi^{1,1}_3=R_1\phi_2\chi^{1,1}_2+R_2\phi_2\chi^{1,1}_4+R_3\phi_1\chi^{1,1}_5$\\
		&$\psi^{1,1}_4=R_1\phi_2\chi^{1,1}_2+R_2\phi_2\chi^{1,1}_5+R_3\phi_1\chi^{1,1}_4$\\
		&$\psi^{1,1}_5=R_1\phi_2\chi^{1,1}_4+R_2\phi_1\chi^{1,1}_2+R_3\phi_1\chi^{1,1}_5$\\
		&$\psi^{1,1}_6=R_1\phi_2\chi^{1,1}_2+R_2\phi_2\chi^{1,1}_4+R_3\phi_2\chi^{1,1}_5+R_4\phi_1\chi^{1,1}_2+R_5\phi_1\chi^{1,1}_4+R_6\phi_1\chi^{1,1}_5$\\
		$2^+$&$\psi^{2,2}_1=R_1\phi_2\chi^{2,2}_1$\\
		&$\psi^{2,2}_2=R_1\phi_1\chi^{2,2}_1$\\
		&$\psi^{2,2}_3=R_1\phi_2\chi^{2,2}_1+R_2\phi_1\chi^{2,2}_1$\\
		\bottomrule[0.5pt]\bottomrule[1.0pt]
	\end{tabular}
\end{table*}

Based on the color SU$(3)_c$ symmetry, one can obtain two combinations of color singlets $6_c\otimes\bar{6}_c$ and $\bar{3}_c\otimes3_c$ for DH tetraquarks, which are
\begin{equation}\label{for:ctetr}
	\begin{aligned}
		\phi_{1}&=\frac{1}{\sqrt{6}}(rr\bar{r}\bar{r}+gg\bar{g}\bar{g}+bb\bar{b}\bar{b})\\
        &+\frac{1}{2\sqrt{6}}(rb\bar{b}\bar{r}+br\bar{b}\bar{r}+gr\bar{g}\bar{r}+rg\bar{g}\bar{r}+gb\bar{b}\bar{g}+bg\bar{b}\bar{g}+gr\bar{g}\bar{r}+rg\bar{r}\bar{g}+gb\bar{g}\bar{b}+bg\bar{g}\bar{b}+rb\bar{r}\bar{b}+br\bar{r}\bar{b}),  \\
        \phi_{2}&=\frac{1}{2\sqrt{3}}(rb\bar{b}\bar{r}-br\bar{b}\bar{r}-gr\bar{g}\bar{r}+rg\bar{g}\bar{r}+gb\bar{b}\bar{g}-bg\bar{b}\bar{g}+gr\bar{r}\bar{g}-rg\bar{r}\bar{g}-gb\bar{g}\bar{b}+bg\bar{g}\bar{b}-rb\bar{r}\bar{b}+br\bar{r}\bar{b}).
	\end{aligned}
\end{equation}

We can combine the flavor, color, and spin wavefunctions, provided that the constraints from the Pauli principle are applied. In the diquark-antidiquark picture, there are 12 possible bases for the wavefunction, having the form of
\begin{equation}
	\begin{aligned}
		&\phi_{1}\chi^{2,2}_{1}=\left|[Q_{1}Q_{2}]_{1}^{6_{c}}[\bar{q}_{3}\bar{q}_{4}]_{1}^{\bar{6}_c}\right\rangle_{2},   		
		&\phi_{2}\chi^{2,2}_{1}=\left|\{Q_{1}Q_{2}\}_{1}^{\bar{3}_{c}}\{\bar{q}_{3}\bar{q}_{4}\}_{1}^{3_{c}}\right\rangle_{2},   	\\
		&\phi_{1}\chi^{1,1}_{2}=\left|[Q_{1}Q_{2}]_{1}^{6_{c}}[\bar{q}_{3}\bar{q}_{4}]_{1}^{\bar{6}_{c}}\right\rangle_{1},   		
		&\phi_{2}\chi^{1,1}_{2}=\left|\{Q_{1}Q_{2}\}_{1}^{\bar{3}_{c}}\{\bar{q}_{3}\bar{q}_{4}\}_{1}^{3_{c}}\right\rangle_{1},   	\\
		&\phi_{1}\chi^{0,0}_{3}=\left|[Q_{1}Q_{2}]_{1}^{6_{c}}[\bar{q}_{3}\bar{q}_{4}]_{1}^{\bar{6}_{c}}\right\rangle_{0},   		
		&\phi_{2}\chi^{0,0}_{3}=\left|\{Q_{1}Q_{2}\}_{1}^{\bar{3}_{c}}\{\bar{q}_{3}\bar{q}_{4}\}_{1}^{3_{c}}\right\rangle_{0},   	\\
		&\phi_{1}\chi^{1,1}_{4}=\left|[Q_{1}Q_{2}]_{1}^{6_{c}}\{\bar{q}_{3}\bar{q}_{4}\}_{0}^{\bar{6}_{c}}\right\rangle_{1},     	
		&\phi_{2}\chi^{1,1}_{4}=\left|\{Q_{1}Q_{2}\}_{1}^{\bar{3}_{c}}[\bar{q}_{3}\bar{q}_{4}]_{0}^{3_{c}}\right\rangle_{1},     	\\
		&\phi_{1}\chi^{1,1}_{5}=\left|\{Q_{1}Q_{2}\}_{0}^{6_{c}}[\bar{q}_{3}\bar{q}_{4}]_{1}^{\bar{6}_{c}}\right\rangle_{1},     	
		&\phi_{2}\chi^{1,1}_{5}=\left|[Q_{1}Q_{2}]_{0}^{\bar{3}_{c}}\{\bar{q}_{3}\bar{q}_{4}\}_{1}^{3_{c}}\right\rangle_{1},     	\\
		&\phi_{1}\chi^{0,0}_{6}=\left|\{Q_{1}Q_{2}\}_{0}^{6_{c}}\{\bar{q}_{3}\bar{q}_{4}\}_{0}^{\bar{6}_{c}}\right\rangle_{0},   	
		&\phi_{2}\chi^{0,0}_{6}=\left|[Q_{1}Q_{2}]_{0}^{\bar{3}_{c}}[\bar{q}_{3}\bar{q}_{4}]_{0}^{3_{c}}\right\rangle_{0},   		\\
	\end{aligned}
\end{equation}
where the symbol $|(\cdots)_{\text{spin}}^{\text{color}}(\cdots)_{\text{spin}}^{\text{color}}\rangle_\text{spin}$ represents the wavefunction with specified color and spin configurations, and the brackets $\{\cdots\}$ and $[\cdots]$ denote the symmetric and antisymmetric diquarks, respectively.
One can combine these basis vectors to construct the wavefunctions of the DH tetraquarks for all possible quantum numbers (as listed in Table \ref{tab:ml}).
In addition, when discussing the decay behavior of these states, the following two sets of wavefunctions are useful (in this work). They consist of the color wavefunctions $8_c\otimes8_c$ and $1_c\otimes1_c$ given by
\begin{equation}\label{for:oc}
	\begin{aligned}
        \phi^\prime_1&=\frac{1}{3\sqrt{2}}(r\bar{r}r\bar{r}+b\bar{b}b\bar{b}+g\bar{g}g\bar{g})+\frac{1}{2\sqrt{2}}(b\bar{r}r\bar{b}+r\bar{b}b\bar{r}+g\bar{r}r\bar{g}+r\bar{g}g\bar{r}+g\bar{b}b\bar{g}+b\bar{g}g\bar{b})\\
		&-\frac{1}{6\sqrt{2}}(r\bar{r}g\bar{g}+g\bar{g}r\bar{r}+b\bar{b}g\bar{g}+g\bar{g}b\bar{b}+b\bar{b}r\bar{r}+r\bar{r}b\bar{b}),\\
        \phi^\prime_2&=\frac{1}{3}(r\bar{r}r\bar{r}+b\bar{b}b\bar{b}+g\bar{g}g\bar{g}+r\bar{r}b\bar{b}+r\bar{r}g\bar{g}+b\bar{b}r\bar{r}+b\bar{b}g\bar{g}+g\bar{g}r\bar{r}+g\bar{g}b\bar{b}),\\
	\end{aligned}
\end{equation}
as well as the three spin wavefunctions
\begin{equation}\label{for:stz}
	\begin{aligned}
        \chi^{1,0}_{2}&=\frac{1}{\sqrt{2}}(\uparrow\uparrow\downarrow\downarrow-\downarrow\downarrow\uparrow\uparrow),\\ \chi^{1,0}_{4}&=\frac{1}{2}(\uparrow\downarrow\uparrow\downarrow+\downarrow\uparrow\uparrow\downarrow-\uparrow\downarrow\downarrow\uparrow-\downarrow\uparrow\downarrow\uparrow),\\ \chi^{1,0}_{5}&=\frac{1}{2}(\uparrow\downarrow\uparrow\downarrow+\uparrow\downarrow\downarrow\uparrow-\downarrow\uparrow\uparrow\downarrow-\downarrow\uparrow\downarrow\uparrow).\\
	\end{aligned}
\end{equation}

 \end{document}